\documentclass[12pt]{article}
\usepackage{amsfonts}
\usepackage{bbm}
\usepackage{bm}
\usepackage{mathrsfs}
\usepackage{color}
\usepackage[left=2.00cm, right=2.00cm,top=2.0cm,bottom=2.0cm]{geometry}             
\usepackage{graphicx}
\usepackage{amssymb}
\usepackage{amsmath}
\usepackage{cite}
\usepackage{comment}
\usepackage{appendix} 
\usepackage{caption}
\usepackage{tikz}
\usetikzlibrary{matrix}
\usepackage{ulem}

\usepackage[hyperindex=true,
          pdfstartview=FitH,
          bookmarksnumbered=true,
          bookmarksopen=true,
          citecolor=blue,
          linkcolor=blue,
          colorlinks=true,
          urlcolor=rossoCP3,
          pdfborder=001,
          unicode]{hyperref}

\usepackage{cancel}
\usepackage{xcolor}  

\allowdisplaybreaks[4]
\definecolor{rossoCP3}{cmyk}{0,.88,.77,.40}
\begin{document}

\title{Covariant kinetic theory of photons coupled to massive particles in curved spacetime}
\author{
  Long Cui${}^1$\thanks{\href{mailto:cuilong@stumail.nwu.edu.cn}{cuilong@stumail.nwu.edu.cn}},~ 
  Yi-fan Cai${}^1$\thanks{\href{mailto:caiyifan@stumail.nwu.edu.cn}{caiyifan@stumail.nwu.edu.cn}},~
  Xin Hao${}^{2,}{}^3$\thanks{\href{mailto:xhao@hebtu.edu.cn}{xhao@hebtu.edu.cn}~(corresponding author)},~ 
  Bin Wu${}^{1,}{}^{4,}{}^{5,}{}^6$\thanks{\href{mailto:binwu@nwu.edu.cn}{binwu@nwu.edu.cn}~(corresponding author)}
\vspace{2pt}\\
\small ${}^1$School of Physics, Northwest University, Xi’an 710127, China\\
\small ${}^2$School of Physics, Hebei Normal University, Shijiazhuang 050024, China\\
\small ${}^3$Shijiazhuang Key Laboratory of Astronomy and Space Science, Shijiazhuang 050024, China\\
\small ${}^4$Shaanxi Key Laboratory for Theoretical Physics Frontiers, Xi’an 710127, China\\
\small ${}^5$Peng Huanwu Center for Fundamental Theory, Xi’an 710127, China\\
\small ${}^6$Fundamental Discipline Research Center for Quantum Science and technology of\\ \small Shaanxi Province, Xi’an 710127, China\\
}

\date{ }

\maketitle

\begin{abstract}
  This paper develops the covariant Boltzmann equation governing the coupled
  evolution of photons and massive particles in general curved spacetimes. We
  construct the collision integrals for photon emission-absorption, Compton
  scattering, and scattering between massive particles, and show the
  conservation laws. Requiring the H-theorem to hold determines the relation
  between the absorption and emission transition rates, which reduces to the Einstein relation in the non-relativistic limit.
  At equilibrium the entropy production vanishes, yielding the detailed
  balance condition, from which we obtain the covariant thermodynamic
  relations of the photon gas. We illustrate these results with the thermal
  radiation from an accretion disk in Kerr spacetime, first in global equilibrium 
  and then in a Keplerian disk in local thermodynamic equilibrium.
  We further extend the collision integrals to arbitrary multi-species
  processes involving both null and massive particles. The formulation
  treats non-equilibrium radiative processes in strong gravitational fields
  from first principles.

\end{abstract}

\section{Introduction}

Radiation is virtually the only messenger from the near-horizon
region of an accreting black hole to a distant observer.
The Event Horizon Telescope's horizon-scale imaging and polarimetry
of M87* and Sgr A* \cite{EventHorizonTelescope:2019dse,EventHorizonTelescope:2021bee,EventHorizonTelescope:2022wkp}
have turned these radiative signatures into directly constrainable observables.
Reliably inferring the state of near-horizon matter from these signals
depends on both the propagation of photons to the detector and the
microscopic processes that generate and release them at the source.
Current models commonly combine general relativistic magnetohydrodynamic (GRMHD)
simulations \cite{Gammie:2003rj} with post-processing ray tracing \cite{Zhou:2025moa}.
This approach has transformed the interpretation of horizon-scale observations,
but it often relies on prescribed electron thermodynamics, assumed particle distributions,
and empirical closures \cite{Moscibrodzka:2015pda,Rowan:2017cao}. Such approximations
can become limiting in dilute, strongly magnetized, and intrinsically non-equilibrium
plasmas near the horizon. Covariant relativistic kinetic theory treats propagation and
collisions within a single framework. The Liouville evolution of the distribution function
describes geometric propagation on the mass shell, while the collision integrals
on the right-hand side of the Boltzmann equation separately encode absorption-emission,
Compton scattering, and more general multi-species reactions. Kinetic theory is therefore
well suited to radiative transport in strong gravitational fields, where the observed signal
traces back through both the curved spacetime and the non-equilibrium interactions that produced
it. Here we develop the framework itself, leaving specific accretion applications to future work.

Kinetic theory reduces the microscopic dynamics to the evolution of distribution functions.
For particles moving near the speed of light, or in strong gravitational fields, this description
must be formulated covariantly. Consequently, relativistic kinetic theory has found wide application
in high-energy physics \cite{Elze:1989un,Arnold:2007pg}, astrophysics \cite{Istomin:2007ge,liu2011nonlinear},
and cosmology \cite{Weinberg:2003ur,vereshchagin2017relativistic}.
The formulation of covariant kinetic theory dates back to the work of Israel \cite{israel1963}, which was subsequently
extended to general curved spacetimes by Stewart \cite{stewart1971} and Ehlers \cite{ehlers1973survey}. In recent years,
Sarbach et al. \cite{Sarbach2012,Sarbach2013,Sarbach2021} reformulated these classical theories within
a rigorous differential geometric setting by introducing the Sasaki metric on the tangent and cotangent bundles.
Despite its mathematical elegance, this geometric construction is restricted to massive particles,
since the Sasaki metric degenerates on the light cone. This degeneracy is not merely a technical issue,
as it renders the standard volume element identically zero for photons. Consequently,
no well-defined covariant Boltzmann equation for photons exists within the existing framework.

In our previous work \cite{cai2025geometry}, the geometric formulation of covariant kinetic theory
is extended to null particles. We introduced a new volume element
on the light cone that satisfies the Liouville theorem and is compatible with the standard
thermodynamic interpretation, thereby providing a consistent geometric description of the
photon gas in curved spacetime. However, the photon gas in realistic astrophysical
environments never evolves in isolation. Photons interact continuously with massive particles,
such as electrons, protons, and ions, through emission, absorption, and Compton scattering.
The coupled evolution of the null and massive sectors drives much of radiative astrophysics
in strong gravity. For such coupled systems, existing approaches cannot treat radiation and
matter on the same footing when the system is driven away from equilibrium by strong gravity
or turbulent dissipation. A unified kinetic description is therefore needed.

The present paper aims to develop a covariant Boltzmann formulation that governs the interactions
between photons and massive particles in general curved spacetimes. This formulation treats
non-equilibrium radiative processes in strong gravitational fields from first principles.
In Section \ref{sec:massive_particles}, we review the geometric formulation of kinetic theory and
its extension from massive to null particles. In Section \ref{sec:Relativistic kinetic theory for null particles},
we derive the collision integrals for photon emission-absorption and Compton scattering,
and prove the corresponding conservation laws and the H-theorem.
In Section \ref{sec:Photon gas in detailed balance state}, we impose the detailed balance condition,
derive the thermodynamic relations of the equilibrium photon gas, and illustrate these results
with the thermal radiation from an equilibrium accretion disk and a Keplerian accretion disk in Kerr spacetime.
In Section \ref{sec:The generalization of collision integral}, we generalize these collision integrals to
arbitrary multi-species reaction processes involving both null and massive particles.
Throughout, we leave the microscopic scattering cross-sections unspecified, keeping the formulation general.

Throughout this paper, the spacetime dimension is taken to be $d+1$, with the metric signature $(-,+,\ldots,+)$.
The Greek indices $\mu,\nu,\ldots = 0,1,\ldots,d$ refer to coordinate indices of the spacetime,
and the Latin indices with hat $\hat{a},\hat{b} = \hat{0},\hat{1},\ldots,\hat{d}$, $\hat{i},\hat{j},\ldots = \hat{1},\hat{2},\ldots,\hat{d}$ refer to frame indices.

\section{Geometric formulation, from massive to null}
\label{sec:massive_particles}

The standard geometric formulation of relativistic kinetic theory for massive particles
breaks down in the massless case. The Sasaki metric, which provides the natural volume
element on the mass shell, becomes degenerate on the light cone. As a result,
the associated volume element that enters the Liouville theorem, the Boltzmann equation,
and the definitions of macroscopic currents vanishes. In this section, we first review
the phase space geometry for massive particles, identifying precisely where the
degeneracy occurs. We then discuss the volume-element dependence of the distribution
function and the entropy flow, which establish the criteria that any alternative
construction must satisfy. Finally, we present a new volume element on the light cone
that satisfies both the Liouville theorem and this criterion.

\subsection{Phase space geometry for massive particles}\label{sec:Phase_space_geometry}

The microscopic state space of a classical particle is defined by its position and momentum. 
Based on~\cite{Sarbach2012,Sarbach2013,Sarbach:2013hna}, we adopt the tangent bundle as the 
phase space. The background spacetime is represented by $M$, and the tangent bundle of $M$ is
\begin{equation}
  T M := \{ (x, p) \mid x \in M,\; p \in T_x M \},
\end{equation}
where $p^\mu = m v^\mu$ is the momentum, $m \neq 0$ the rest mass, and $v^\mu$ the proper 
velocity of the massive particle; $T_x M$ denotes the tangent space at $x$. 

To construct a metric on $TM$ that behaves covariantly under coordinate transformations 
on the base manifold, one cannot simply use $\mathrm{d}p^\mu$ for the fiber direction, since 
it does not transform as a tensor. Instead, one introduces the vertical 1-form
\begin{equation}
  \theta^\mu = \mathrm{d}p^\mu + \Gamma^\mu_{\alpha\nu} p^\alpha \mathrm{d}x^\nu,
\end{equation}
where the Christoffel term compensates the non-tensorial part of $\mathrm{d}p^\mu$, 
ensuring that $\theta^\mu$ transforms in the same way as $\mathrm{d}x^\mu$. The natural 
metric on the tangent bundle $TM$, known as the Sasaki metric~\cite{sasaki1958differential}, 
is then given by
\begin{align}
  \hat{g} = g_{\mu\nu}\,\mathrm{d}x^\mu \otimes \mathrm{d}x^\nu 
         + g_{\mu\nu}\,\theta^\mu \otimes \theta^\nu.
\end{align}
In the basis \(\{\mathrm{d}x^\mu, \theta^\mu\}\), the Sasaki metric is block-diagonal,
and each block has determinant \(g = |\det(g_{\mu\nu})|\). The associated volume
element is therefore
\begin{equation}
  \boldsymbol{\eta}_{TM} = g \, \mathrm{d}x^0 \wedge \cdots \wedge \mathrm{d}x^d 
                        \wedge \theta^0 \wedge \cdots \wedge \theta^d
                        = g \, \mathrm{d}x^0 \wedge \cdots \wedge \mathrm{d}x^d 
                        \wedge \mathrm{d}p^0 \wedge \cdots \wedge \mathrm{d}p^d,
\end{equation}
where the equality follows because the connection terms in \(\theta^\mu\)
cancel in the full antisymmetrized product.

In the absence of scattering, the phase trajectory of a charged particle in the
gravitational and electromagnetic fields follows the geodesic equation supplemented by
the Lorentz force, and its tangent vector is the Liouville vector field \cite{Sarbach:2013hna}
\begin{equation}\label{Liouville vector field}
  L = p^{\mu}  \cfrac{\partial}{\partial x^{\mu}} + \left(q F^\mu {}_\nu p^\nu - \Gamma^{\mu}_{\alpha \beta} p^{\alpha} p^{\beta} \right) 
  \cfrac{\partial}{\partial p^{\mu}},
\end{equation}
where $q$ represents the amount of charge of the particle and $F_{\mu \nu}$ is the
field strength tensor of the electromagnetic field. It is straightforward to verify
that the mass of the particle is a conserved quantity
\begin{align}
  L(g_{\mu\nu}p^\mu p^\nu)=0.
\end{align}
Consequently, the conservation of this quantity along the phase trajectory confines the
microscopic state to the mass shell
\begin{equation}
  \Gamma_m := \{ (x, p) \in T M|p^{\mu} p^{\nu} g_{\mu \nu} = - m^2 c^2 \}.
\end{equation}
Let us denote the fiber space in $\Gamma_m$ as $P_m:=T_x M\cap \Gamma_m$.
The corresponding volume
element compatible with the induced metric of the Sasaki metric on
$\Gamma_m$ can be attained through the
contraction of $\ensuremath{\boldsymbol{\eta}}_{T M}$ and the unit normal
vector of the mass shell, yielding
\begin{equation}
  \ensuremath{\boldsymbol{\eta}}_{\Gamma_m} = g \frac{m c}{| p_0 |} \mathrm{d}
  x^0 \wedge \cdots \wedge \mathrm{d} x^d \wedge \mathrm{d} p^1 \wedge \cdots
  \wedge \mathrm{d} p^d = \ensuremath{\boldsymbol{\eta}}_M \wedge
  \ensuremath{\boldsymbol{\eta}}_{P_m}, \label{eq:volume Gamma m}
\end{equation}
where
\begin{equation}
  \ensuremath{\boldsymbol{\eta}}_M = \sqrt{g} \, \mathrm{d} x^0 \wedge \cdots
  \wedge \mathrm{d} x^d
\end{equation}
is the volume element of background spacetime $M$ and
\begin{equation}
  \ensuremath{\boldsymbol{\eta}}_{P_m} = \sqrt{g} \, \frac{m c}{| p_0 |}
  \mathrm{d} p^1 \wedge \cdots \wedge \mathrm{d} p^d
\end{equation}
is the volume element of fiber space $P_m$.

Another well-known Lorentz-invariant volume element on $P_m$ is
\begin{align}
  \ensuremath{\boldsymbol{\varpi}} = \frac{1}{m c}
  \ensuremath{\boldsymbol{\eta}}_{P_m} = \sqrt{g} \frac{1}{| p_0 |} \mathrm{d}
p^1 \wedge \cdots \wedge \mathrm{d} p^d.
\end{align}
Note that the factor $m$ cancels in $\ensuremath{\boldsymbol{\varpi}}$,
which will prove {instructive} when we construct the null-particle volume element.
Beyond this, this volume element can simplify the expressions of thermodynamic
quantities that will appear later. Therefore, we primarily use it in the following
paragraphs.

\subsection{Liouville evolution and the Boltzmann equation}

The distribution function associated with the volume element
\eqref{eq:volume Gamma m} is denoted by \(f\). Its physical meaning
follows from the Lie derivative
\begin{align}\label{eq:meaning-f}
  \mathcal{L}_L (f \boldsymbol{\eta}_{\Gamma_m}),
\end{align}
which represents the rate of change of particle number per unit phase volume along
the phase trajectory. When the particle undergoes scattering, 
it jumps from one phase trajectory to another, causing a change in the particle density
around the original trajectory. This change is encoded by the scattering integral
\begin{equation}\label{eq:scattering-integral-o}
  \mathcal{C} (x, p_1) = \int_{P_m \times P_m \times P_m} \boldsymbol{\varpi}_2 
  \boldsymbol{\varpi}_3  \boldsymbol{\varpi}_4  \left[W
  (p_3 + p_4 \to p_1 + p_2) f_3 f_4 - W (p_1 + p_2
  \to p_3 + p_4) f_1 f_2\right],
\end{equation}
so that
\begin{align}
  \mathcal{L}_L (f\boldsymbol{\eta}_{\Gamma_m})=\mathcal{C}\boldsymbol{\eta}_{\Gamma_m}.
\end{align}
Using Liouville theorem on the mass shell, \(\mathcal{L}_L\boldsymbol{\eta}_{\Gamma_m}=0\),
this reduces to the standard Boltzmann equation
\begin{align}
  L[f]=\mathcal{C}.\label{eq:boltzmann-equation}
\end{align}

Particle flow $N^{\mu}$ and energy-momentum tensor $T^{\mu \nu}$ are
important physical quantities to describe fluid properties. From an intuitive
standpoint, the particle flow represents the distribution-weighted integral
of the single-particle proper velocity over the fiber space $P_m$
\begin{equation}
  N^{\mu} = \int_{P_m} \ensuremath{\boldsymbol{\eta}}_{P_m} f v^{\mu} = c \int_{P_m}
  \ensuremath{\boldsymbol{\varpi}} f p^{\mu}, \label{eq:N1}
\end{equation}
and the energy-momentum tensor $T^{\mu \nu}$ is the distribution-weighted integral of
the single-particle momentum flow over the fiber space $P_m$
\begin{equation}
  T^{\mu \nu} = \int_{P_m} \ensuremath{\boldsymbol{\eta}}_{P_m} f p^{\mu} v^{\nu} =
  c \int_{P_m} \ensuremath{\boldsymbol{\varpi}} f p^{\mu} p^{\nu}. \label{eq:T1}
\end{equation}
Eqs. \eqref{eq:N1}-\eqref{eq:T1} can also be rigorously derived from the interpretation of
Eq.~\eqref{eq:meaning-f}, as detailed in \cite{cai2025geometry}.

For a degenerate quantum system, the collision integral receives a statistical
correction from the occupation number of the final state. To treat Bose, Fermi,
and Boltzmann statistics in a unified manner, the transition intensity is multiplied
by the quantum correction factor $f^{\ast} = 1 + \varsigma \mathfrak{g}^{-1} h^d f$,
where $\varsigma = 0, 1, -1$ for the non-degenerate, Bose-Einstein,
and Fermi-Dirac cases respectively, $\mathfrak{g}$ is the degeneracy factor,
and $h$ is the Planck constant. {A self-contained derivation of these quantum
statistical factors is provided in Appendix \ref{sec:quantum_statistical_factors}}.
The quantum-corrected collision integral then reads
\begin{equation}\label{eq:scattering-integral}
  \mathcal{C} (x, p_1) = \int \ensuremath{\boldsymbol{\varpi}}_2
  \ensuremath{\boldsymbol{\varpi}}_3  \ensuremath{\boldsymbol{\varpi}}_4  [W
  (p_3 + p_4 \to p_1 + p_2) f_3 f_4 f_1^{\ast} f_2^{\ast} - W (p_1 + p_2
  \to p_3 + p_4) f_1 f_2 f_3^{\ast} f_4^{\ast}],
\end{equation}
where we have, for simplicity, omitted the explicit notation of the integration domain,
a convention we follow for all fiber-space integrations in what follows.

Throughout this section the collision integral is written for massive particles.
In Section \ref{sec:Relativistic kinetic theory for null particles},
the same structure, with the appropriate phase-space measure and transition rates,
will be specialized to processes involving photons, including absorption-emission
and Compton scattering.

\subsection{Volume-element dependence and the entropy flow} \label{sec:Volume-element dependence and the entropy flow}

The interpretation of the distribution function \( f \) based on Eq.~\eqref{eq:meaning-f}
suggests that \( f \) depends on the choice of the volume element of the mass shell bundle.
When we select a different volume element
\begin{align}
  \ensuremath{\boldsymbol{\eta}}_{\Gamma_m} \to k \ensuremath{\boldsymbol{\eta}}_{\Gamma_m},
\end{align}
the distribution function \( f \) also changes accordingly
\begin{align}
  f \to k^{-1}f,\label{eq:f-change}
\end{align}
so that the physical combination \( f \boldsymbol{\eta}_{\Gamma_m} \) remains unchanged.
Since \(\boldsymbol{\eta}_{\Gamma_m}\) decomposes as
\(m c \, \boldsymbol{\eta}_M \wedge \boldsymbol{\varpi}\), and \(\boldsymbol{\eta}_M\)
is unchanged under this rescaling, the product \(\boldsymbol{\varpi} f\) is also invariant.
Because the definitions \eqref{eq:N1} and \eqref{eq:T1} involve only the combination
\(\boldsymbol{\varpi} f\), the particle flow and the energy-momentum tensor are independent
of the choice of the volume element.

Inspired by ensemble theory of equilibrium states, the entropy flow vector \(S^{\mu}\)
is usually defined as (see~\cite{cercignani2002})
\begin{equation}
  S^{\mu} = - k_B c \int
  \boldsymbol{\varpi} p^{\mu} f \left[ \log \left( \frac{h^d
  f}{f^{\ast} \mathfrak{g}} \right) - \frac{\mathfrak{g} \log
  f^{\ast}}{\varsigma h^d f} \right], \label{eq:entropy flow}
\end{equation}
where \(k_B\) is the Boltzmann constant and \(h^d\) ensures that the argument of
the logarithm is dimensionless\footnote{For \(\varsigma=0\), the expression is
understood in the limit \(\varsigma\to0\), where it remains finite.}.
Since entropy is a physical quantity, it cannot depend on the arbitrary
choice of the volume element. Therefore, the combination \(h^d f\) must remain
invariant under the rescaling of \(\boldsymbol{\eta}_{\Gamma_m}\). Consequently,
the Planck constant must transform as
\begin{align}\label{eq:h-change}
  h\to k^{1/d}  h.
\end{align}
It is worth emphasizing that the transformed Planck constant should be
regarded only as a dimension-balancing constant, since it appears solely through
the invariant combination \(h^d f\). Every integrated observable therefore retains
the standard Planck constant. This freedom must be kept under control once the
volume element can no longer be inherited from a metric, as happens on the
light cone.

\subsection{The null-particle challenge and its resolution}

We now return to the geometric obstruction identified in Section
\ref{sec:Phase_space_geometry}. For null particles, the phase space
is the light cone
\[
  {\Gamma_0 := \{ (x, p) \in T M \mid p^{\mu} p^{\nu} g_{\mu\nu} = 0 \},}
\]
and the fiber space in $\Gamma_0$ is denoted as $P_0 := T_x M \cap \Gamma_0$.
The Sasaki metric becomes degenerate on $\Gamma_0$, and consequently the induced
volume element \eqref{eq:volume Gamma m} vanishes in the massless limit because
it contains an overall factor of \(m\). Every structure built on this volume
element collapses with it. The Lie derivative
$\mathcal{L}_L(f\ensuremath{\boldsymbol{\eta}}_{\Gamma_m})$, the Boltzmann equation
\eqref{eq:boltzmann-equation}, and the definitions of $N^\mu$, $T^{\mu\nu}$,
and $S^\mu$ as fiber integrals over $P_m$ all lose their geometric foundation.
One cannot build a kinetic theory on a zero volume element.

An alternative volume element $\ensuremath{\boldsymbol{\eta}}_{\Gamma_0}$ must be
supplied. Guided by the structure of the massive-particle theory, we require that
$\ensuremath{\boldsymbol{\eta}}_{\Gamma_0}$ satisfy two criteria:
\begin{enumerate}
   \item The Liouville theorem $\mathcal{L}_L \ensuremath{\boldsymbol{\eta}}_{\Gamma_0}=0$
   must hold, so that the Boltzmann equation retains the simple form \eqref{eq:boltzmann-equation};
   \item The combination \(h^d f\) must retain its dimensional normalization in the
   entropy flow, as required by the volume-element independence argument of
   Section \ref{sec:Volume-element dependence and the entropy flow}.
 \end{enumerate}
These criteria naturally extend the massive-particle formulation to the null case.
They are not unique in a purely logical sense, but they are the minimal requirements
that preserve the standard kinetic structure and prevent the proliferation of spurious
normalization factors in the thermodynamic quantities.

A volume element on the light cone that satisfies both criteria is
\begin{align}\label{eq:eta-Gamma-0}
  \ensuremath{\boldsymbol{\eta}}_{\Gamma_0} := c\ensuremath{\boldsymbol{\eta}}_{M} \wedge \ensuremath{\boldsymbol{\varpi}} = g \frac{c}{| p_0 |} \mathrm{d} x^0 \wedge \cdots \wedge \mathrm{d} x^d \wedge \mathrm{d} p^1 \wedge \cdots \wedge \mathrm{d} p^d.
\end{align}
The first criterion is automatically satisfied by construction $\mathcal{L}_L \ensuremath{\boldsymbol{\eta}}_{\Gamma_0} =\mathcal{L}_L
\ensuremath{\boldsymbol{\eta}}_{\Gamma_m} / m = 0$. The second criterion follows from the definition of the entropy flow. On the light cone, the entropy flow is defined by the same expression as in the massive case  \eqref{eq:entropy flow}, which involves only \(\ensuremath{\boldsymbol{\varpi}} f\) and \(h^d f\), and both remain well defined in the massless limit. The combination \(h^d f\) therefore keeps the dimensional normalization assumed by this definition. Once the phase-space measure on the light cone is fixed in this way, the kinetic theory of photons can be developed in the next section.


\section{Relativistic kinetic theory for photons}
\label{sec:Relativistic kinetic theory for null particles}
In an accreting black hole system, the astrophysical plasma is a mixture of photons and
massive particles, and its electromagnetic radiation is the primary window to the
near-horizon region. The microscopic processes that shape this radiation involve both
photons and massive particles. Photon-photon scattering, however, is far too weak to
thermalize the photon gas on its own\footnote{Other processes such as pair production
\(\gamma\gamma\to e^+e^-\) also exist but are beyond the scope of this work.},
so the relevant interactions are those between photons and massive particles,
including absorption, emission, and Compton scattering. For simplicity,
this section considers a system composed of photons and multiple species of
neutral massive particles, interacting through absorption, emission and Compton
scattering together with scattering between massive particles, while reaction processes
between massive particles are neglected\footnote{We use ``reaction processes''
to refer to a process in which the species of particles is changed, such as a chemical
reaction.}.

For clarity, subscript `$\gamma$' denotes quantities exclusive to photons,
and subscript `$A$' denotes quantities associated with massive particles.
When no subscript appears, the discussion applies universally to both. For example,
the distribution function of photons is $f_\gamma$, and that of a massive particle
$A$ is $f_A$. The Lorentz-invariant volume element on the momentum space is
$\ensuremath{\boldsymbol{\varpi}}_\gamma$ for photons and
$\ensuremath{\boldsymbol{\varpi}}_A$ for massive particles. The Liouville vector
fields are
\begin{equation}
  L_A = p_A^{\mu}  \cfrac{\partial}{\partial x^{\mu}} - \Gamma^{\mu}_{\alpha
  \beta} p_A^{\alpha} p_A^{\beta}  \cfrac{\partial}{\partial p_A^{\mu}} ,
\end{equation}
\begin{equation}
  L_\gamma = p_\gamma^{\mu}  \cfrac{\partial}{\partial x^{\mu}} - \Gamma^{\mu}_{\alpha
  \beta} p_\gamma^{\alpha} p_\gamma^{\beta}  \cfrac{\partial}{\partial p_\gamma^{\mu}} .
\end{equation}

Generally, the Boltzmann equation for any component in the system can be written as
\begin{equation}
  L [f] = \sum_i \mathcal{C}_i (x, p) . \label{eq:Boltzmann equation i}
\end{equation}
On the left-hand side is the derivative of the distribution function along the Liouville vector field,
signifying that particles move on the mass shell. On the right-hand side is the collision integral.
A particle typically participates in various collision processes, and we therefore sum over all processes
involving that particle. In the present system, we have
\begin{equation}
  L_\gamma [f_\gamma] =\mathcal{C}_{\text{rad}} (x, p_\gamma) +\mathcal{C}_{\text{com}} (x, p_\gamma).
  \label{eq:Boltzmann 0}
\end{equation}
for photons, and
\begin{equation}
  L_A [f_A] =\mathcal{C}_{\text{rad}} (x, p_A) +\mathcal{C}_{\text{com}} (x, p_A) +\mathcal{C}_{\text{sca}}
  (x, p_A). \label{eq:Boltzmann 1}
\end{equation}
for each massive particle species $A$, where $\mathcal{C}_{\text{rad}}$
collectively denotes the absorption-emission processes, $\mathcal{C}_{\text{com}}$
denotes Compton scattering, and $\mathcal{C}_{\text{sca}}$ denotes scattering between
massive particles.

To construct these collision integrals from transition rates, we consider an arbitrary
region $\hat{V} \subset \Gamma_m$ bounded by the integral curves of the Liouville vector
field, lying between two $2d$-dimensional spacelike hypersurfaces
$\hat{\Sigma}_1, \hat{\Sigma}_2$ (representing two spatial cross-sections at the
given times $t_1, t_2$) on the mass shell. The regions associated with photons and with
massive particle $A$ are denoted $\hat{V}_\gamma$ and $\hat{V}_A$, respectively.

The variation in the number of particle trajectories within region $\hat{V} \subset \Gamma_m$ is given by
\begin{equation}
  \Delta N = \int_{\hat{V}} \ensuremath{\boldsymbol{\eta_M}} \wedge
  \ensuremath{\boldsymbol{\varpi}} \, L [f] . \label{eq:DN}
\end{equation}
The derivation of equation \eqref{eq:DN} is provided in the Appendix
\ref{sec:Derivation DN}. Substituting \eqref{eq:Boltzmann 0} and
\eqref{eq:Boltzmann 1} into \eqref{eq:DN}, we obtain
\begin{align}
  \Delta N_\gamma &= \Delta_{\text{rad}} N_\gamma + \Delta_{\text{com}} N_\gamma, \label{eq:DN-gamma}\\
  \Delta N_A &= \Delta_{\text{rad}} N_A + \Delta_{\text{com}} N_A + \Delta_{\text{sca}} N_A, \label{eq:DN-A}
\end{align}
where, for instance, $\Delta_{\text{rad}} N_\gamma =
\displaystyle{\int_{\hat{V}_\gamma} \ensuremath{\boldsymbol{\eta_M}} \wedge
\ensuremath{\boldsymbol{\varpi}}_\gamma \, \mathcal{C}_{\text{rad}}}$, and
similarly for the remaining terms. These contributions can be expressed
directly in terms of the transition rates. Equating the two forms yields the
collision integrals, as
demonstrated below.

\subsection{Absorption and emission of photons}\label{sec:absorption emission}
For the process of photon absorption and emission, \(p_\gamma + p_A' \leftrightarrow p_A\),
where particles are labeled by their momenta, a massive particle \(A\) transitions between
energy levels by absorbing or emitting a photon. Although the rest mass differs slightly
between the initial and final states due to internal energy changes, this difference
is negligible for our purposes.\footnote{We also neglect differences in level degeneracy
throughout this section for simplicity.} More complex processes such as bremsstrahlung
are not treated here; a general framework for multi-species reactions is presented
in Section \ref{sec:The generalization of collision integral},
leaving specific cross-sections unspecified.

For photon absorption, we introduce the transition rate $W_{\text{abs}} (p_\gamma +
p_A' \to p_A)$, where subscript `$\text{abs}$' represents absorption. Under the
constraint of momentum conservation, $W_{\text{abs}}$ should satisfy the condition
\begin{equation}
  W_{\text{abs}} (p_\gamma + p_A' \to p_A) \propto \delta^{d + 1} \left( p^{\mu}_\gamma {+
  p^{\mu}_A}' - p^{\mu}_A \right) . \label{eq:Wa}
\end{equation}
This transition rate $W_{\text{abs}} (p_\gamma + p_A' \to p_A)$ effectively measures the
total number of photon trajectories absorbed by massive particles within
$\hat{V}_\gamma \subset \Gamma_0$
\begin{equation}
  \Delta_{\text{abs}} N_\gamma = - \sum_A \int_{\hat{V}_\gamma} \ensuremath{\boldsymbol{\eta}}_M
  \wedge \ensuremath{\boldsymbol{\varpi}}_\gamma \int
  \ensuremath{\boldsymbol{\varpi}}_A \wedge
  \ensuremath{\boldsymbol{\varpi}}_A' W_{\text{abs}} (p_\gamma + p_A' \to p_A) f_\gamma f_A'
  f_A^{\ast}, \label{eq:DaN0}
\end{equation}
where the minus sign accounts for the loss of photons due to absorption, and the sum over
\(A\) runs over all massive species.

To avoid repeatedly writing products of transition rates and distribution
functions, we introduce the statistical transition kernel for a
collision process with incoming and outgoing particles,
\begin{equation}
  \mathcal{W}_{(\text{in} \to \text{out})} :=
  W(\text{in} \to \text{out}) \prod_{i \in \text{in}} f_i \prod_{j \in \text{out}} f^*_j .
\end{equation}
The subscript specifies the incoming and outgoing particle species.
When the physical nature of the process needs to be distinguished
(absorption, emission, Compton, etc.), a superscript is added. For example,
the absorption kernel defined above is
\begin{equation}
  \mathcal{W}^{\text{abs}}_{(\gamma + A' \to A)} :=
  W_{\text{abs}}(p_\gamma + p_A' \to p_A) f_\gamma f_A' f_A^*,
\end{equation}
and Eq.~\eqref{eq:DaN0} simplifies to
\begin{equation}
  \Delta_{\text{abs}} N_\gamma = - \sum_A \int_{\hat{V}_\gamma}
  \boldsymbol{\eta}_M \wedge \boldsymbol{\varpi}_\gamma \int
  \boldsymbol{\varpi}_A \wedge \boldsymbol{\varpi}_A' \;
  \mathcal{W}^{\text{~abs}}_{(\gamma + A' \to A)}.
\end{equation}
This compact notation will be used to express the collision integrals.
Whenever the explicit products are needed, as in the derivations of the conservation laws
and the H-theorem below, we restore the full form.

For massive particle $A$, the forward
collision $p_\gamma + p_A' \to p_A$ causes particle $A$ to enter the region
$\hat{V}_A \subset \Gamma_{m_A}$
\[ \int_{\hat{V}_A} \boldsymbol{\eta}_M \wedge \boldsymbol{\varpi}_A
\int \boldsymbol{\varpi}_\gamma \wedge \boldsymbol{\varpi}_A' \;
\mathcal{W}^{\text{~abs}}_{(\gamma + A' \to A)} . \]
Whereas the reverse collision $p_\gamma + p_A \to p_A'$ causes particle $A$ to
leave the region $\hat{V}_A \subset \Gamma_{m_A}$
\[ \int_{\hat{V}_A} \boldsymbol{\eta}_M \wedge \boldsymbol{\varpi}_A
\int \boldsymbol{\varpi}_\gamma \wedge \boldsymbol{\varpi}_A' \;
\mathcal{W}^{\text{~abs}}_{(\gamma + A \to A')} . \]
Consequently, for the absorption process, the net change in the total number of
particle $A$ trajectories within $\hat{V}_A$ is

\begin{equation}
   \Delta_{\text{abs}} N_A = \int_{\hat{V}_A} \boldsymbol{\eta}_M \wedge
  \boldsymbol{\varpi}_A \int \boldsymbol{\varpi}_\gamma \wedge
  \boldsymbol{\varpi}_A' \; [\mathcal{W}^{\text{~abs}}_{(\gamma + A' \to A)}
  - \mathcal{W}^{\text{~abs}}_{(\gamma + A \to A')}] .
  \label{eq:DaNA}
\end{equation}

For photon emission, introducing the transition rate
$W_{\text{emi}} (p_A \to p_A' + p_\gamma)$, where the subscript `$\text{emi}$'
represents emission.
Due to momentum conservation, $W_{\text{emi}}$ satisfies
\begin{equation}
  W_{\text{emi}} (p_A \to p_A' + p_\gamma) \propto \delta^{d + 1} \left( p^{\mu}_A
  {- p^{\mu}_A}' - p^{\mu}_\gamma \right) . \label{eq:We}
\end{equation}
The increment in photon trajectories within the region $\hat{V}_\gamma \subset
\Gamma_0$ is

\begin{equation}
   \Delta_{\text{emi}} N_\gamma = \sum_A \int_{\hat{V}_\gamma} \boldsymbol{\eta}_M
   \wedge \boldsymbol{\varpi}_\gamma \int \boldsymbol{\varpi}_A 
   \wedge \boldsymbol{\varpi}_A' \; \mathcal{W}^{\text{~emi}}_{(A \to A' + \gamma)} .
\label{eq:DeN0}
\end{equation}

Similar to equation \eqref{eq:DaNA}, for the emission process, the net change in
the total number of particle $A$ trajectories within $\hat{V}_A \subset
\Gamma_{m_A}$ is
\begin{equation}
  \Delta_{\text{emi}} N_A = \int_{\hat{V}_A} \boldsymbol{\eta}_M
  \wedge \boldsymbol{\varpi}_A \int \boldsymbol{\varpi}_\gamma
  \wedge \boldsymbol{\varpi}_A' \; [\mathcal{W}^{\text{~emi}}_{(A' \to A + \gamma)}
  - \mathcal{W}^{\text{~emi}}_{(A \to A' + \gamma)}] .
  \label{eq:DeNA}
\end{equation}

Therefore, combining equations \eqref{eq:DaN0} and \eqref{eq:DeN0}, the change
in the number of photon trajectories in region $\hat{V}_\gamma$ due to the
absorption-emission process is
\begin{align*}
  \Delta_{\text{rad}} N_\gamma  = &~ \Delta_{\text{abs}} N_\gamma + \Delta_{\text{emi}} N_\gamma\\
  = &~ \sum_A \int_{\hat{V}_\gamma} \boldsymbol{\eta}_M \wedge \boldsymbol{\varpi}_\gamma
  \int \boldsymbol{\varpi}_A \wedge \boldsymbol{\varpi}_A' \; [\mathcal{W}^{\text{~emi}}_{(A \to A' + \gamma)}
  - \mathcal{W}^{\text{~abs}}_{(\gamma + A' \to A)}].
\end{align*}

According to \eqref{eq:DN-gamma} and the arbitrariness of
$\hat{V}_\gamma$, the collision integral for photons from the
absorption-emission process is
\begin{equation}
  \mathcal{C}_{\text{rad}} (x, p_\gamma) = \sum_A \int \boldsymbol{\varpi}_A
  \wedge \boldsymbol{\varpi}_A' \; [\mathcal{W}^{\text{~emi}}_{(A \to A' + \gamma)}
  - \mathcal{W}^{\text{~abs}}_{(\gamma + A' \to A)}] .
  \label{eq:Cr0}
\end{equation}

Similarly, for massive particles, the change in the number of trajectories of
massive particle $A$ in region $\hat{V}_A$ due to the absorption-emission
process is
\begin{align*}
  \Delta_{\text{rad}} N_A  = &~ \Delta_{\text{abs}} N_A + \Delta_{\text{emi}} N_A\\
   = &~ \int_{\hat{V}_A} \boldsymbol{\eta}_M \wedge \boldsymbol{\varpi}_A
   \int \boldsymbol{\varpi}_\gamma \wedge \boldsymbol{\varpi}_A'\;
   \bigl( [\mathcal{W}^{\text{~abs}}_{(\gamma + A' \to A)} - \mathcal{W}^{\text{~abs}}_{(\gamma + A \to A')}]
   + [\mathcal{W}^{\text{~emi}}_{(A' \to A + \gamma)} - \mathcal{W}^{\text{~emi}}_{(A \to A' + \gamma)}] \bigr) .
\end{align*}
Due to the arbitrariness of region $\hat{V}_A$, for the Boltzmann equation of
massive particle $A$, the collision integral contributed by the
absorption-emission process is
\begin{align}
  \mathcal{C}_{\text{rad}} (x, p_A) = \int \boldsymbol{\varpi}_\gamma
  \wedge \boldsymbol{\varpi}_A' \; \bigl( [\mathcal{W}^{\text{~abs}}_{(\gamma + A' \to A)}
  - \mathcal{W}^{\text{~abs}}_{(\gamma + A \to A')}] + [\mathcal{W}^{\text{~emi}}_{(A' \to A + \gamma)}
  - \mathcal{W}^{\text{~emi}}_{(A \to A' + \gamma)}] \bigr) .
  \label{eq:CrA}
\end{align}

\subsection{Compton scattering and scattering of massive
particles}\label{sec:scattering}

In addition to photon absorption and emission, we also consider Compton
scattering and scattering of massive particles themselves, while photon
self-scattering is neglected.

For Compton scattering, $p_\gamma + p_A \to p_\gamma' + p_A'$, introducing the
corresponding transition rate $W_{\text{com}} (p_\gamma + p_A \to p_\gamma' + p_A')$, with the
subscript `$\text{com}$' denoting Compton scattering. According to the conservation of
momentum,
\begin{equation}
  W_{\text{com}} (p_\gamma + p_A \to p_\gamma' + p_A') \propto \delta^{d + 1} \left( p^{\mu}_\gamma
  + p^{\mu}_A {- p^{\mu}_\gamma}' {- p^{\mu}_A}' \right) . \label{eq:Wc}
\end{equation}
Taking into account the alteration in the number of particle trajectories in
phase space before and after scattering, the forward collision $p_\gamma' + p_A'
\to p_\gamma + p_A$ inject photons into the region $\hat{V}_\gamma \subset \Gamma_0$
\[ \sum_A \int_{\hat{V}_\gamma} \boldsymbol{\eta}_M \wedge \boldsymbol{\varpi}_\gamma \int \boldsymbol{\varpi}_A \wedge \boldsymbol{\varpi}_\gamma' \wedge \boldsymbol{\varpi}_A' \; \mathcal{W}^{\text{~com}}_{(\gamma' + A' \to \gamma + A)} , \]
whereas the reverse collision ejects $p_\gamma + p_A \to p_\gamma' + p_A'$ particles
from $\hat{V}_\gamma$
\[ \sum_A \int_{\hat{V}_\gamma} \boldsymbol{\eta}_M \wedge \boldsymbol{\varpi}_\gamma \int \boldsymbol{\varpi}_A \wedge \boldsymbol{\varpi}_\gamma' \wedge \boldsymbol{\varpi}_A' \; \mathcal{W}^{\text{~com}}_{(\gamma + A \to \gamma' + A')} . \]
Thus, in the case of photons, the modification to the trajectory count within
$\hat{V}_\gamma$ is given by

\begin{equation}
  \Delta_{\text{com}} N_\gamma = \sum_A \int_{\hat{V}_\gamma} \boldsymbol{\eta}_M \wedge \boldsymbol{\varpi}_\gamma \int \boldsymbol{\varpi}_A \wedge \boldsymbol{\varpi}_\gamma' \wedge \boldsymbol{\varpi}_A' \; [\mathcal{W}^{\text{~com}}_{(\gamma' + A' \to \gamma + A)} - \mathcal{W}^{\text{~com}}_{(\gamma + A \to \gamma' + A')}] .
  \label{eq:DcN0}
\end{equation}

Likewise, for a massive particle $A$, the variation in its trajectory count
within $\hat{V}_A \subset \Gamma_{m_A}$ is quantified by

\begin{equation}
  \Delta_{\text{com}} N_A = \int_{\hat{V}_A} \boldsymbol{\eta}_M \wedge \boldsymbol{\varpi}_A \int \boldsymbol{\varpi}_\gamma \wedge \boldsymbol{\varpi}_\gamma' \wedge \boldsymbol{\varpi}_A' \; [\mathcal{W}^{\text{~com}}_{(\gamma' + A' \to \gamma + A)} - \mathcal{W}^{\text{~com}}_{(\gamma + A \to \gamma' + A')}] .
  \label{eq:DcNA}
\end{equation}

Therefore, from \eqref{eq:DN-gamma}--\eqref{eq:DN-A} and the arbitrariness of
$\hat{V}_\gamma$ and $\hat{V}_A$, we obtain the contributions of Compton
scattering to the collision integrals for photons and massive particles,
respectively

\begin{equation}
  \mathcal{C}_{\text{com}} (x, p_\gamma) = \sum_A \int \boldsymbol{\varpi}_A \wedge \boldsymbol{\varpi}_\gamma' \wedge \boldsymbol{\varpi}_A' \; [\mathcal{W}^{\text{~com}}_{(\gamma' + A' \to \gamma + A)} - \mathcal{W}^{\text{~com}}_{(\gamma + A \to \gamma' + A')}] .
  \label{eq:Cc0}
\end{equation}

\begin{equation}
  \mathcal{C}_{\text{com}} (x, p_A) = \int \boldsymbol{\varpi}_\gamma \wedge \boldsymbol{\varpi}_\gamma' \wedge \boldsymbol{\varpi}_A' \; [\mathcal{W}^{\text{~com}}_{(\gamma' + A' \to \gamma + A)} - \mathcal{W}^{\text{~com}}_{(\gamma + A \to \gamma' + A')}] .
  \label{eq:CcA}
\end{equation}

We now consider scattering between massive particles, $p_A + p_B \to p_A'
+ p_B'$, for which reaction processes are disregarded for now. We introduce the
transition rate $W_{\text{sca}} (p_A + p_B \to p_A' + p_B')$, where the subscript `$\text{sca}$'
represents the scattering between massive particles. The momentum conservation
dictates
\begin{equation}
  W_{\text{sca}} (p_A + p_B \to p_A' + p_B') \propto \delta^{d + 1} \left( p^{\mu}_A
  + p^{\mu}_B {- p^{\mu}_A}' {- p^{\mu}_B}' \right) . \label{eq:Ws}
\end{equation}
By subtracting the effects of reverse collisions from those of forward
collisions, we can ascertain the net change in the number of particle $A$
trajectories within region $\hat{V}_A \subset \Gamma_{m_A}$, which equates
to

\begin{equation}
  \Delta_{\text{sca}} N_A = \sum_B \int_{\hat{V}_A} \boldsymbol{\eta}_M \wedge \boldsymbol{\varpi}_A \int \boldsymbol{\varpi}_B \wedge \boldsymbol{\varpi}_A' \wedge \boldsymbol{\varpi}_B' \; [\mathcal{W}^{\text{~sca}}_{(A' + B' \to A + B)} - \mathcal{W}^{\text{~sca}}_{(A + B \to A' + B')}] .
  \label{eq:DsNA}
\end{equation}

Thus, the contribution to the collision integral for massive particles due to
scattering between massive particles is

\begin{equation}
  \mathcal{C}_{\text{sca}} (x, p_A) = \sum_B \int \boldsymbol{\varpi}_B \wedge \boldsymbol{\varpi}_A' \wedge \boldsymbol{\varpi}_B' \; [\mathcal{W}^{\text{~sca}}_{(A' + B' \to A + B)} - \mathcal{W}^{\text{~sca}}_{(A + B \to A' + B')}] .
  \label{eq:CsA}
\end{equation}
The Boltzmann equations \eqref{eq:Boltzmann 0} and \eqref{eq:Boltzmann 1} are now established, with each collision integral written in terms of a generic transition rate. To apply the formalism to realistic astrophysical plasmas, these rates must be further specified from the underlying microscopic physics. 

Before doing so, however, we first derive two general results that follow from the structure of the collision integrals themselves, namely the conservation laws and the H-theorem. The former ensures that particle number and energy-momentum are conserved at the macroscopic level, while the latter guarantees non-negative entropy production, consistent with the second law of thermodynamics. These results are therefore central to connecting the kinetic description to relativistic hydrodynamics and thermodynamics, and their derivation is presented below.

\subsection{Conservation law}
The flow defined below, originally introduced for massive particles,
can be extended to null particles~\cite{cai2025geometry}
\begin{equation}
  \mathcal{O}^{\mu} = \int \ensuremath{\boldsymbol{\varpi}} p^{\mu}
  \mathcal{O}f, \label{eq:flowO}
\end{equation}
and its covariant divergence can be expressed as
\begin{equation}
  \nabla_{\mu} \mathcal{O}^{\mu} = \int \ensuremath{\boldsymbol{\varpi}}
  L [\mathcal{O}f] . \label{eq:divO}
\end{equation}
It can also be attained by employing a similar derivation of equation
\eqref{eq:DN} (see Appendix \ref{sec:Derivation DN}), wherein $f$ is replaced
by $\mathcal{O}f$, and invoking Gauss theorem.

Considering $\mathcal{O}= c$, equation \eqref{eq:divO} yields
\begin{equation}
  \nabla_{\mu} N^{\mu} = c \int \ensuremath{\boldsymbol{\varpi}} L [f] .
  \label{eq:divN}
\end{equation}
Taking $\mathcal{O}= c p^{\mu} X_{\mu}$, with $X_{\mu}$ being an arbitrary
covector field in spacetime. By using Eqs.~\eqref{eq:T1}, \eqref{eq:flowO} and \eqref{eq:divO}, a direct computation shows that
\[ X_{\nu} \nabla_{\mu} T^{\mu \nu} + T^{\mu \nu} \nabla_{\mu} X_{\nu} = c
   \int \ensuremath{\boldsymbol{\varpi}}  \left( p^{\mu} X_{\mu} L [f] +
   p^{\mu} p^{\nu} \nabla_{\mu} X_{\nu} f + q F^{\mu}_{\enspace \nu} p^{\nu}
   X_{\mu} f \right) . \]
According to the arbitrariness of $X_{\mu}$, it follows that
\begin{equation}
  \nabla_{\mu} T^{\mu \nu} = q F^{\mu}_{\enspace \nu} N^{\nu} + c \int
  \ensuremath{\boldsymbol{\varpi}} p^{\nu} L [f] . \label{eq:divT}
\end{equation}
Equations \eqref{eq:divN} and \eqref{eq:divT} are well-known formulations in
kinetic theory, universally applicable to both massive particles and null
particles.

Under our assumptions, photons are subject to absorption and emission,
so their number is not conserved. Massive particles, by contrast, undergo no
chemical reactions, so their individual species numbers are conserved.
The total energy-momentum tensor, including both photons and massive particles,
is expected to be conserved. We now verify these statements from the collision
integrals derived above.

Let us first examine the conservation of the number of particles. For massive particles, we have
\begin{align}
  \nabla_{\mu} N_A^{\mu} \nonumber
  = &~ c \int \ensuremath{\boldsymbol{\varpi}}_A  (\mathcal{C}_{\text{rad}} (x, p_A)
  +\mathcal{C}_{\text{com}} (x, p_A) +\mathcal{C}_{\text{sca}} (x, p_A)) \nonumber\\
   = &~ c \int \ensuremath{\boldsymbol{\varpi}}_A
  \ensuremath{\boldsymbol{\varpi}}_\gamma \ensuremath{\boldsymbol{\varpi}}_A'
  \; \bigl( [\mathcal{W}^{\text{~abs}}_{(\gamma + A' \to A)} - \mathcal{W}^{\text{~abs}}_{(\gamma + A \to A')}]
  + [\mathcal{W}^{\text{~emi}}_{(A' \to A + \gamma)} - \mathcal{W}^{\text{~emi}}_{(A \to A' + \gamma)}] \bigr) \nonumber\\
    &~ + c \int \ensuremath{\boldsymbol{\varpi}}_A
  \ensuremath{\boldsymbol{\varpi}}_\gamma \ensuremath{\boldsymbol{\varpi}}_\gamma'
  \ensuremath{\boldsymbol{\varpi}}_A'
  \; [\mathcal{W}^{\text{~com}}_{(\gamma' + A' \to \gamma + A)} - \mathcal{W}^{\text{~com}}_{(\gamma + A \to \gamma' + A')}] \nonumber\\
    &~ + c \sum_B \int \ensuremath{\boldsymbol{\varpi}}_A
  \ensuremath{\boldsymbol{\varpi}}_B \ensuremath{\boldsymbol{\varpi}}_A'
  \ensuremath{\boldsymbol{\varpi}}_B'
  \; [\mathcal{W}^{\text{~sca}}_{(A' + B' \to A + B)} - \mathcal{W}^{\text{~sca}}_{(A + B \to A' + B')}] \nonumber\\
   = &~ 0.
\end{align}
The last step exploits the exchange symmetry $A \leftrightarrow A'$, $\gamma
\leftrightarrow \gamma'$, $B \leftrightarrow B'$. For instance, expanding the
scattering kernel,
\[
  \mathcal{W}^{\text{~sca}}_{(A' + B' \to A + B)} = W_{\text{sca}}(p_{A'} + p_{B'} \to p_A + p_B)
  f_{A'} f_{B'} f_A^{\ast} f_B^{\ast},
\]
the exchange $A \leftrightarrow A'$, $B \leftrightarrow B'$ maps it to
\[
  W_{\text{sca}}(p_A + p_B \to p_{A'} + p_{B'}) f_A f_B f_{A'}^{\ast} f_{B'}^{\ast}
  = \mathcal{W}^{\text{~sca}}_{(A + B \to A' + B')}.
\]
The exchange therefore reverses the sign of each bracket, so its integral vanishes.
The radiative and Compton brackets vanish for the same reason, so $\nabla_\mu N_A^\mu = 0$.

Similarly, for photons,
using equation \eqref{eq:divN} and Boltzmann equation \eqref{eq:Boltzmann 0},
yields
\begin{align*}
  \nabla_{\mu} N_\gamma^{\mu}
   = &~ c \int \ensuremath{\boldsymbol{\varpi}}_\gamma L_\gamma [f_\gamma]
   = c \int \ensuremath{\boldsymbol{\varpi}}_\gamma (\mathcal{C}_{\text{rad}} (x, p_\gamma)
  +\mathcal{C}_{\text{com}} (x, p_\gamma))\\
   = &~ c \sum_A \int \ensuremath{\boldsymbol{\varpi}}_\gamma
  \ensuremath{\boldsymbol{\varpi}}_A \ensuremath{\boldsymbol{\varpi}}_A'
  \; [\mathcal{W}^{\text{~emi}}_{(A \to A' + \gamma)}
  - \mathcal{W}^{\text{~abs}}_{(\gamma + A' \to A)}]\\
    &~ + c \sum_A \int \ensuremath{\boldsymbol{\varpi}}_\gamma
  \ensuremath{\boldsymbol{\varpi}}_A \ensuremath{\boldsymbol{\varpi}}_\gamma'
  \ensuremath{\boldsymbol{\varpi}}_A'
  \; [\mathcal{W}^{\text{~com}}_{(\gamma' + A' \to \gamma + A)}
  - \mathcal{W}^{\text{~com}}_{(\gamma + A \to \gamma' + A')}] .
\end{align*}
The Compton term vanishes by the exchange symmetry $\gamma \leftrightarrow \gamma'$, $A \leftrightarrow A'$.
The radiative term does not vanish, because absorption removes a photon and emission creates one,
so the forward and reverse processes are not related by any exchange of momenta.
Photon number is therefore not conserved.

For the total energy-momentum tensor, employing equation \eqref{eq:divT},
Boltzmann equation \eqref{eq:Boltzmann 0}, \eqref{eq:Boltzmann 1}, and the
exchange symmetry $A \leftrightarrow A', \gamma \leftrightarrow \gamma', B
\leftrightarrow B'$, we derive that
\begin{align}
    &~ \nabla_{\mu} \left( T_\gamma^{\mu \nu} + \sum_A T^{\mu \nu}_A \right)
  \nonumber\\
   = &~ c \int \ensuremath{\boldsymbol{\varpi}}_\gamma (\mathcal{C}_{\text{rad}} (x, p_\gamma)
  +\mathcal{C}_{\text{com}} (x, p_\gamma)) p_\gamma^{\nu} + c \sum_A \int
  \ensuremath{\boldsymbol{\varpi}}_A  (\mathcal{C}_{\text{rad}} (x, p_A) +\mathcal{C}_{\text{com}
  } (x, p_A) +\mathcal{C}_{\text{sca}} (x, p_A)) p_A^{\nu} \nonumber\\
   = &~ \frac{c}{2} \sum_A \int \ensuremath{\boldsymbol{\varpi}}_\gamma
  \ensuremath{\boldsymbol{\varpi}}_A \ensuremath{\boldsymbol{\varpi}}_A'
  \Big[ \bigl(\mathcal{W}^{\text{~emi}}_{(A \to A' + \gamma)} - \mathcal{W}^{\text{~abs}}_{(\gamma + A' \to A)}\bigr)
  \left( p_\gamma^{\nu} {+ p_A^{\nu}}' - p_A^{\nu} \right) \nonumber \\
  &~ \qquad \qquad \qquad \qquad \quad + \bigl(\mathcal{W}^{\text{~emi}}_{(A' \to A + \gamma)} - \mathcal{W}^{\text{~abs}}_{(\gamma + A \to A')}\bigr) \left( p_\gamma^{\nu} + p_A^{\nu} {- p_A^{\nu}}' \right) \Big] \nonumber\\
    &~ + \frac{c}{2} \sum_A \int \ensuremath{\boldsymbol{\varpi}}_\gamma
  \ensuremath{\boldsymbol{\varpi}}_A \ensuremath{\boldsymbol{\varpi}}_\gamma'
  \ensuremath{\boldsymbol{\varpi}}_A'
  \bigl[\mathcal{W}^{\text{~com}}_{(\gamma' + A' \to \gamma + A)} - \mathcal{W}^{\text{~com}}_{(\gamma + A \to \gamma' + A')}\bigr]
  \left( p_\gamma^{\nu} + p_A^{\nu} {- p_\gamma^{\nu}}' {- p_A^{\nu}}' \right) \nonumber\\
    &~ + \frac{c}{4} \sum_{A, B} \int \ensuremath{\boldsymbol{\varpi}}_A
  \ensuremath{\boldsymbol{\varpi}}_B \ensuremath{\boldsymbol{\varpi}}_A'
  \ensuremath{\boldsymbol{\varpi}}_B'
  \bigl[\mathcal{W}^{\text{~sca}}_{(A' + B' \to A + B)} - \mathcal{W}^{\text{~sca}}_{(A + B \to A' + B')}\bigr]
  \left( p_A^{\nu} + p_B^{\nu} {- p_A^{\nu}}' {- p_B^{\nu}}' \right) \nonumber\\
   = &~ 0. 
\end{align}
where the factors $1/2$ and $1/4$ arise from symmetrizing over the forward and reverse processes.
In the last step we make use of the momentum conservation condition, see equations
\eqref{eq:Wa}, \eqref{eq:We}, \eqref{eq:Wc} and \eqref{eq:Ws}.  
In short, although photon number is not conserved in the coupled system,
particle number for each massive species and the total energy-momentum tensor
are still conserved.

\subsection{H-theorem}\label{sec:H-theorem}
We now turn to the H-theorem, which is the statistical-mechanical form of the second law of
thermodynamics. Without microscopic reversibility and number conservation, it is natural to
ask whether the H-theorem still holds, and what it then requires of the absorption and
emission rates.

By invoking equation \eqref{eq:divO} and setting 
\[\mathcal{O}_\gamma= - k_B \left(
\log \left( \frac{h^d f_\gamma}{f_\gamma^{\ast}   \mathfrak{g}_\gamma} \right) -
\frac{\mathfrak{g}_\gamma \log f_\gamma^{\ast}}{\varsigma_\gamma h^d f_\gamma}
\right) \footnote{For photons, $\varsigma_\gamma = 1, \mathfrak{g}_\gamma = d - 1$, yet we
retain the form of $\varsigma_\gamma$ and $\mathfrak{g}_\gamma$ in subsequent
calculation to ensure the generality of the computation process.}, ~~
\mathcal{O}_A= - k_B \left( \log \left( \frac{h^d f_A}{f_A^{\ast}  
\mathfrak{g}_A} \right) - \frac{\mathfrak{g}_A \log f_A^{\ast}}{\varsigma_A
h^d f_A} \right)\]
respectively, using exchange symmetry $A \leftrightarrow A',
\gamma \leftrightarrow \gamma', B \leftrightarrow B'$, we calculate the divergence of
the total entropy of the system as
\begin{align}
    &~ \nabla_{\mu} S^{\mu}_\gamma + \sum_A \nabla_{\mu} S^{\mu}_A \nonumber\\
   = &~ c \int \ensuremath{\boldsymbol{\varpi}}_\gamma L_\gamma \left[ - k_B f_\gamma 
  \left( \log \left( \frac{h^d f_\gamma}{f_\gamma^{\ast}   \mathfrak{g}_\gamma}
  \right) - \frac{\mathfrak{g}_\gamma \log f_\gamma^{\ast}}{\varsigma_\gamma h^d f_\gamma}
  \right) \right] \nonumber\\
    & + \sum_A c \int \ensuremath{\boldsymbol{\varpi}}_A L_A \left[ - k_B
  f_A  \left( \log \left( \frac{h^d f_A}{f_A^{\ast}   \mathfrak{g}_A} \right)
  - \frac{\mathfrak{g}_A \log f_A^{\ast}}{\varsigma_A h^d f_A} \right) \right]
  \nonumber\\
   = & - k_B c \int \ensuremath{\boldsymbol{\varpi}}_\gamma \log
  \frac{h^d f_\gamma}{\mathfrak{g}_\gamma f_\gamma^{\ast}}  (\mathcal{C}_{\text{rad}} (x, p_\gamma)
  +\mathcal{C}_{\text{com}} (x, p_\gamma)) \nonumber\\
    & - k_B c \int \ensuremath{\boldsymbol{\varpi}}_A \log \frac{h^d
  f_A}{\mathfrak{g}_A f_A^{\ast}}  (\mathcal{C}_{\text{rad}} (x, p_A) +\mathcal{C}_{\text{com}}
  (x, p_A) +\mathcal{C}_{\text{sca}} (x, p_A)) \nonumber\\
   = &~ (\nabla_{\mu} S^{\mu})_{\text{sca}} + (\nabla_{\mu} S^{\mu})_{\text{com}}  + (\nabla_{\mu}
  S^{\mu})_{\text{rad}}, 
\end{align}
where we denote respective contributions to entropy generation from Compton
scattering, scatterings between massive particles, and radiative process as
\begin{align}
(\nabla_{\mu} S^{\mu})_{\text{sca}} & = - k_B c \int \boldsymbol{\varpi}_A \log \frac{h^d f_A}{\mathfrak{g}_A f_A^{\ast}} \mathcal{C}_{\text{sca}} (x, p_A) \nonumber\\
& = \frac{k_B c}{4} \sum_{A, B} \int \boldsymbol{\varpi}_A \boldsymbol{\varpi}_{B} \boldsymbol{\varpi}_A' \boldsymbol{\varpi}_B' \; \log \frac{f_A f_B {f_A^{\ast}}' {f_B^{\ast}}'}{f_A' f_B' f_A^{\ast} f_B^{\ast}} \; [\mathcal{W}^{\text{~sca}}_{(A,B \to A',B')} - \mathcal{W}^{\text{~sca}}_{(A',B' \to A,B)}] , \label{eq:dSs} \\
(\nabla_{\mu} S^{\mu})_{\text{com}} & = - k_B c \int \boldsymbol{\varpi}_\gamma \log \frac{h^d f_\gamma}{\mathfrak{g}_\gamma f_\gamma^{\ast}} \mathcal{C}_{\text{com}} (x, p_\gamma) - k_B c \int \boldsymbol{\varpi}_A \log \frac{h^d f_A}{\mathfrak{g}_A f_A^{\ast}} \mathcal{C}_{\text{com}} (x, p_A) \nonumber\\
& = \frac{k_B c}{2} \sum_A \int \boldsymbol{\varpi}_\gamma \boldsymbol{\varpi}_A \boldsymbol{\varpi}_\gamma' \boldsymbol{\varpi}_A' \; \log \frac{f_\gamma f_A {f_\gamma^{\ast}}' {f_A^{\ast}}'}{f_\gamma' f_A' f_\gamma^{\ast} f_A^{\ast}} \; [\mathcal{W}^{\text{~com}}_{(\gamma + A \to \gamma' + A')} - \mathcal{W}^{\text{~com}}_{(\gamma' + A' \to \gamma + A)}] , \label{eq:dSc} \\
(\nabla_{\mu} S^{\mu})_{\text{rad}} & = - k_B c \int \boldsymbol{\varpi}_\gamma \log \frac{h^d f_\gamma}{\mathfrak{g}_\gamma f_\gamma^{\ast}} \mathcal{C}_{\text{rad}} (x, p_\gamma) - k_B c \int \boldsymbol{\varpi}_A \log \frac{h^d f_A}{\mathfrak{g}_A f_A^{\ast}} \mathcal{C}_{\text{rad}} (x, p_A) \nonumber\\
& = k_B c \sum_A \int \boldsymbol{\varpi}_\gamma \boldsymbol{\varpi}_A \boldsymbol{\varpi}_A' \; \log \frac{h^d f_\gamma f_A' f_A^{\ast}}{\mathfrak{g}_\gamma f_\gamma^{\ast} f_A {f_A^{\ast}}'} \; [\mathcal{W}^{\text{~abs}}_{(\gamma + A' \to A)} - \mathcal{W}^{\text{~emi}}_{(A \to A' + \gamma)}] . \label{eq:dSr}
\end{align}

These three entropy productions originate from distinct physical processes.
To ensure that any system strictly adheres to the second law of thermodynamics
(i.e., H-theorem $\nabla_{\mu} S^{\mu} \geqslant 0$), we need to require that the
entropy production from each physical process is no less than zero, i.e.,
$(\nabla_{\mu} S^{\mu})_{\text{sca}} \geqslant 0, (\nabla_{\mu} S^{\mu})_{\text{com}}
\geqslant 0, (\nabla_{\mu} S^{\mu})_{\text{rad}} \geqslant 0$.

For the $(\nabla_{\mu} S^{\mu})_{\text{sca}}$ and $(\nabla_{\mu} S^{\mu})_{\text{com}}$ terms, noting the
microscopic reversibility of binary scattering
\begin{equation}
  W_{\text{sca}} (p_A + p_B \to p_A' + p_B') = W_{\text{sca}} (p_A' + p_B' \to p_A + p_B),
\end{equation}
\begin{equation}
  W_{\text{com}} (p_\gamma + p_A \to p_\gamma' + p_A') = W_{\text{com}} (p_\gamma' + p_A' \to p_\gamma + p_A),
\end{equation}
the $W$ factor cancels between the numerator and denominator of each logarithm
in \eqref{eq:dSs} and \eqref{eq:dSc}, and the entropy productions reduce to
\begin{align*}
  (\nabla_{\mu} S^{\mu})_{\text{sca}} & = \frac{k_B c}{4} \sum_{A, B} \int \boldsymbol{\varpi}_A \boldsymbol{\varpi}_{B} \boldsymbol{\varpi}_A' \boldsymbol{\varpi}_B' \; \log \frac{\mathcal{W}^{\text{~sca}}_{(A,B \to A',B')}}{\mathcal{W}^{\text{~sca}}_{(A',B' \to A,B)}} \; [\mathcal{W}^{\text{~sca}}_{(A,B \to A',B')} - \mathcal{W}^{\text{~sca}}_{(A',B' \to A,B)}] , \\
  (\nabla_{\mu} S^{\mu})_{\text{com}} & = \frac{k_B c}{2} \sum_A \int \boldsymbol{\varpi}_\gamma \boldsymbol{\varpi}_A \boldsymbol{\varpi}_\gamma' \boldsymbol{\varpi}_A' \; \log \frac{\mathcal{W}^{\text{~com}}_{(\gamma + A \to \gamma' + A')}}{\mathcal{W}^{\text{~com}}_{(\gamma' + A' \to \gamma + A)}} \; [\mathcal{W}^{\text{~com}}_{(\gamma + A \to \gamma' + A')} - \mathcal{W}^{\text{~com}}_{(\gamma' + A' \to \gamma + A)}] .
\end{align*}
Applying the inequality \((x - y)\log(x/y)\geqslant0\) with \(x\) and \(y\) as the
forward and reverse transition kernels gives \((\nabla_\mu S^\mu)_{\text{sca}} \geqslant 0\)
and \((\nabla_\mu S^\mu)_{\text{com}} \geqslant 0\).

For the radiative part \eqref{eq:dSr}, absorption and emission change the photon number,
so there is no microscopic reversibility of the binary-scattering type to invoke.
In this situation, requiring $(\nabla_\mu S^\mu)_{\text{rad}} \geqslant 0$ determines the
relation between the two rates:
\begin{equation}
  W_{\text{abs}} (p_\gamma + p_A' \to p_A) = \frac{h^d}{\mathfrak{g}_\gamma}
  W_{\text{emi}} (p_A \to p_A' + p_\gamma), \label{eq:Wa,We}
\end{equation}
which reduces to the Einstein relation governing radiation processes \cite{einstein20167}
in the non-relativistic limit, see Appendix \ref{sec:einstein_coefficients} for details.
Assuming this relation, the radiative contribution can be shown to satisfy
$(\nabla_\mu S^\mu)_{\text{rad}} \geqslant 0$. Thus, the system satisfies
\begin{equation}
  \nabla_\mu S^\mu_\gamma + \sum_A \nabla_\mu S^\mu_A
  = (\nabla_\mu S^\mu)_{\text{com}} + (\nabla_\mu S^\mu)_{\text{sca}} + (\nabla_\mu S^\mu)_{\text{rad}} \geqslant 0
\end{equation}
if and only if the Einstein relation \eqref{eq:Wa,We} holds.


\section{Photon gas in detailed balance state}\label{sec:Photon gas in
detailed balance state}

\subsection{The detailed balance condition}

When a system reaches equilibrium, the entropy production of the system becomes
zero, in which case equations \eqref{eq:dSc}, \eqref{eq:dSs} and \eqref{eq:dSr}
are zero. Hence yielding
\begin{equation}
  \frac{f_\gamma f_A {f_\gamma^{\ast}}' {f_A^{\ast}}'}{f_\gamma' f_A' f_\gamma^{\ast} f_A^{\ast}}
  = \frac{f_A f_B {f_A^{\ast}}' {f_B^{\ast}}'}{f_A' f_B' f_A^{\ast}
  f_B^{\ast}} = \frac{h^d f_\gamma f_A' f_A^{\ast}}{\mathfrak{g}_\gamma
  f_\gamma^{\ast} f_A {f_A^{\ast}}'} = 1.
\end{equation}
Equivalently, in logarithmic form
\begin{eqnarray*}
  \log \frac{h^d f_\gamma}{\mathfrak{g}_\gamma f_\gamma^{\ast}} + \log \frac{h^d
  f_A}{\mathfrak{g}_A f_A^{\ast}} = \log
  \frac{h^d f_\gamma'}{\mathfrak{g}_\gamma {f_\gamma^{\ast}}'} + \log \frac{h^d
  f_A'}{\mathfrak{g}_A {f_A^{\ast}}'}, &  & \text{(Compton scattering)}\\
  \log \frac{h^d f_A}{\mathfrak{g}_A f_A^{\ast}} + \log \frac{h^d
  f_B}{\mathfrak{g}_B f_B^{\ast}} = \log \frac{h^d f_A'}{\mathfrak{g}_A
  {f_A^{\ast}}'} + \log \frac{h^d f_B'}{\mathfrak{g}_B {f_B^{\ast}}'}, &  &
  \text{(Scatterings between massive particles)}\\
  \log \frac{h^d f_\gamma}{\mathfrak{g}_\gamma f_\gamma^{\ast}} + \log \frac{h^d
  f_A'}{\mathfrak{g}_A {f_A^{\ast}}'} = \log \frac{h^d f_A}{\mathfrak{g}_A
  f_A^{\ast}} . &  & \text{(Absorption and emission)}
\end{eqnarray*}
The two sides of these three equations represent the conserved quantities before and after each collision. In equilibrium, the logarithmic combination for each species must be a linear combination of the conserved quantities. The relevant conserved quantities are the total momentum, shared by all processes, and, for each massive species, the particle number and an associated scalar \(\alpha_A\), both invariant under collisions. Thus, for massive particles, the logarithmic combination depends on both scalar and momentum; for photons, only the momentum term is allowed. To ensure that all three collision equations hold simultaneously, the momentum coefficient must be the same across all species. Hence we have
\begin{equation}
  \log \dfrac{h^d f_\gamma}{\mathfrak{g}_\gamma f_\gamma^{\ast}} = \mathcal{B}_{\mu}
  p_\gamma^{\mu}, \quad \log \frac{h^d f_A}{\mathfrak{g}_A f_A^{\ast}} = \mathcal{B}_{\mu}
  p_A^{\mu} - \alpha_A,
\end{equation}
where \(\mathcal{B}_\mu\) is the common covector introduced above. Inverting these relations gives

\begin{equation}
  f_\gamma = \frac{\mathfrak{g}_\gamma}{h^d }
  \frac{1}{\mathrm{e}^{-\mathcal{B}_{\mu} p_\gamma^{\mu}} - \varsigma_\gamma}, \quad f_A
  = \frac{\mathfrak{g}_A}{h^d} \frac{1}{\mathrm{e}^{\alpha_A -\mathcal{B}_{\mu}
  p_A^{\mu}} - \varsigma_A} . \label{eq:detailed balance f}
\end{equation}

When the entropy production rate vanishes, the above conditions imply that the
collision integral vanishes. This means that the equilibrium distribution function
must satisfy the Boltzmann equation with a vanishing collision integral.
Substituting \eqref{eq:detailed balance f} into
\[ L_\gamma [f_\gamma] = 0, \quad L_A [f_A] = 0, \]
we obtain the detailed balance condition.
\begin{equation}
  \nabla_{\mu} \alpha_A = 0, \hspace{1em} \quad \nabla_{(\mu}
  \mathcal{B}_{\nu)} = 0.
\end{equation}
Thus, in global equilibrium, \(\alpha_A\) is a constant scalar and \(\mathcal{B}^{\mu}\) is a Killing vector field. In the non-relativistic limit, \(\mathcal{B}_{\mu} p^{\mu}\) must reduce to the particle energy, which requires \(\mathcal{B}^{\mu}\) to be timelike. A global equilibrium state can therefore exist only in stationary spacetimes.

The detailed balance distribution function \eqref{eq:detailed balance f} determines the
thermodynamic quantities in equilibrium. The physical meanings of $\alpha_A$ and $\mathcal{B}^{\mu}$ 
follow from the relations between these thermodynamic quantities \cite{Hao:2021ifw}.
\begin{equation}
  \alpha_A = - \frac{\mu_A}{k_B T}, \quad \mathcal{B}^{\mu} = \beta U^{\mu} =
  \frac{1}{k_B T} U^{\mu},
\end{equation}
where $\mu_A$ is the chemical potential for massive particle $A$, $T$ is the
temperature for both photon and massive particle $A$, $U^{\mu}$ is the proper
velocity of the fluid system, and we introduce $\beta = 1 / (k_B T)$ for the
convenience of subsequent expression. 

It is worth noting that in detailed balance, the photon gas and the massive-particle gas have the same temperature and proper velocity. Although individual photons follow null worldlines, the photon gas as a whole has a timelike proper velocity, defined as the average of the photon momenta. Since photons propagate in all directions, this average is timelike. Thus the photon gas can be assigned a rest frame, just like the massive-particle gas. In this common rest frame, the photon gas appears as blackbody radiation, while the massive particles follow their respective equilibrium distributions. Continuous absorption and emission ensure that the two components remain at rest relative to each other.

\subsection{Thermodynamic relations of the photon gas}

The equilibrium particle flow and energy-momentum tensor for the photon gas have been computed in our companion paper \cite{Wang:2026kce} using the orthonormal frame and momentum-space integrals. The results are
\begin{equation}
  N^\mu = \frac{\mathfrak{g}_\gamma}{h^d}\frac{\mathcal{A}_{d-1}}{c^d\beta^d}\mathcal{I}_{d-1}(\varsigma_\gamma) U^\mu,
  \label{eq:N eq0}
\end{equation}
\begin{equation}
  T^{\mu\nu} = \frac{\mathfrak{g}_\gamma}{h^d}\frac{\mathcal{A}_{d-1}}{c^d\beta^{d+1}}\mathcal{I}_d(\varsigma_\gamma)
  \left(\frac{1}{c^2}U^\mu U^\nu + \frac{1}{d}\Delta^{\mu\nu}\right),
  \label{eq:T eq0}
\end{equation}
where $\mathcal{A}_{d-1}$ is the area of the unit sphere $S^{d-1}$ and $\Delta^{\mu\nu} = g^{\mu\nu} + U^\mu U^\nu/c^2$ is the projection tensor. We also use the special function
\begin{equation}
  \mathcal{I}_n(\varsigma) := \int_0^\infty \frac{x^n \mathrm{d}x}{e^x-\varsigma}
  = \varsigma^{-1}\Gamma(n+1)\operatorname{Li}_{n+1}(\varsigma), \qquad
  \operatorname{Li}_n(\varsigma) = \sum_{k=1}^\infty \frac{\varsigma^k}{k^n},
\end{equation}
which is the same integral that appears in \cite{Wang:2026kce}. The particle number density, energy density, and pressure measured by a comoving observer are
\begin{equation}
  n = \frac{\mathfrak{g}_\gamma}{h^d}\frac{\mathcal{A}_{d-1}}{c^d\beta^d}\mathcal{I}_{d-1}(\varsigma_\gamma), \quad
  \epsilon = \frac{\mathfrak{g}_\gamma}{h^d}\frac{\mathcal{A}_{d-1}}{c^d\beta^{d+1}}\mathcal{I}_d(\varsigma_\gamma), \quad
  P = \frac{1}{d}\epsilon.
\end{equation}
The results are consistent with those obtained from non-relativistic equilibrium thermodynamics.
Relativistic effects manifest in observer transformations, and we will show the thermodynamic
relations of the equilibrium photon gas measured by a generic observer.

Now let us derive the thermodynamic relations of the photon gas in equilibrium
state. First, we calculate the entropy flux of the equilibrium photon gas.
\begin{eqnarray}
  S^{\mu} & = & - k_B c \int \ensuremath{\boldsymbol{\varpi}}_\gamma \, p_\gamma^{\mu} f _\gamma
  \left[ \log \left( \frac{h^d f_\gamma }{f^{\ast}_\gamma   \mathfrak{g}_\gamma} \right) -
  \frac{{\mathfrak{g}_\gamma \log f^{\ast}_\gamma }}{\varsigma_\gamma h^d f_\gamma} \right]
  \nonumber\\
  & = & - k_B c \int \ensuremath{\boldsymbol{\varpi}}_\gamma \, p_\gamma^{\mu}  \left[
  \frac{\mathfrak{g}_\gamma}{h^d} \frac{1}{\mathrm{e}^{- p_\gamma^{\mu} \mathcal{B}_{\mu}}
  - \varsigma_\gamma} \mathcal{B}_{\nu} p_\gamma^{\nu} - \frac{\mathfrak{g}_\gamma}{\varsigma_\gamma
  h^d} \log \left( \frac{1}{1 - \mathrm{e}^{p_\gamma^{\mu} \mathcal{B}_{\mu}}
  \varsigma_\gamma} \right) \right] \nonumber\\
  & = & - k_B  \left[ \mathcal{B}_{\nu} T^{\mu \nu} - c \int
  \ensuremath{\boldsymbol{\varpi}}_\gamma p_\gamma^{\mu} \frac{\mathfrak{g}_\gamma}{\varsigma_\gamma
  h^d} \log \left( \frac{1}{1 - \mathrm{e}^{p_\gamma^{\mu} \mathcal{B}_{\mu}}
  \varsigma_\gamma} \right) \right] \nonumber\\
  & = & k_B (-\mathcal{B}_{\nu} T^{\mu \nu} + P\mathcal{B}^{\mu}),
  \label{eq:S eq0} 
\end{eqnarray}
where in the final step, we perform integration by parts on the second term,
yielding
\[ c \int \ensuremath{\boldsymbol{\varpi}}_\gamma \, p_\gamma^{\mu}
   \frac{\mathfrak{g}_\gamma}{\varsigma_\gamma h^d} \log \left( \frac{1}{1 -
   \mathrm{e}^{p_\gamma^{\mu} \mathcal{B}_{\mu}} \varsigma_\gamma} \right) =\mathcal{A}_{d
   - 1}  \frac{\mathfrak{g}_\gamma}{\varsigma_\gamma h^d}  \frac{1}{d} U^{\mu} \int
   \mathrm{d} p^{\hat{0}} (p^{\hat{0}})^d  \frac{c \beta
   \varsigma_\gamma}{\mathrm{e}^{c \beta p^{\hat{0}}} - \varsigma_\gamma} = \beta P
   U^{\mu} . \]
Equation \eqref{eq:S eq0} represents the covariant Euler relation of the photon
gas in equilibrium state.

From the intermediate step in \eqref{eq:S eq0}, we identify
\begin{equation}
  P\mathcal{B}^{\mu} = c \int \ensuremath{\boldsymbol{\varpi}}_\gamma p_\gamma^{\mu}
  \frac{\mathfrak{g}_\gamma}{\varsigma_\gamma h^d} \log \left( \frac{1}{1 -
  \mathrm{e}^{p_\gamma^{\mu} \mathcal{B}_{\mu}} \varsigma_\gamma} \right) .
\end{equation}
It is easy to verify that
\[ \frac{\partial (P\mathcal{B}^{\mu})}{\partial \mathcal{B}_{\nu}} = c \int
   \ensuremath{\boldsymbol{\varpi}}_\gamma p_\gamma^{\mu} p_\gamma^{\nu}
   \frac{\mathfrak{g}_\gamma}{h^d}  \frac{1}{\mathrm{e}^{- p_\gamma^{\mu}
   \mathcal{B}_{\mu}} - \varsigma_\gamma} = T^{\mu \nu} . \]
From this, we obtain the covariant Gibbs-Duhem relation for the photon gas in
equilibrium state
\begin{equation}
  \mathrm{d} (P\mathcal{B}^{\mu}) = T^{\mu \nu} \mathrm{d} \mathcal{B}_{\nu} .
  \label{eq:Gibbs-Duhem}
\end{equation}
Combining equations \eqref{eq:S eq0} and \eqref{eq:Gibbs-Duhem}, we obtain the
covariant first law of thermodynamics for the equilibrium photon gas
\begin{equation}
  \mathrm{d} S^{\mu} = k_B (-\mathcal{B}_{\nu} \mathrm{d} T^{\mu \nu}) .
  \label{eq:first law}
\end{equation}
Although the calculations in this subsection are restricted to global equilibrium, 
the key input was the form of the equilibrium distribution function, not the full detailed balance condition.
Thus for a photon gas system that is slightly out of
equilibrium but admits a local equilibrium description, the
hydrodynamic variables at the local equilibrium order still satisfy the
thermodynamic relations \eqref{eq:S eq0}, \eqref{eq:Gibbs-Duhem}, and
\eqref{eq:first law}.

Let us now discuss the observer transformations. We introduce a generic observer \(\mathscr{Z}\) with proper velocity \(Z^\mu\). The proper velocity of the equilibrium photon gas can be decomposed as
\begin{equation}
  U^{\mu} = \gamma (Z^{\mu} + u^{\mu}),
\end{equation}
where \(\gamma = -c^{-2} Z_\mu U^\mu\) is the Lorentz factor and \(u^{\mu}\) is the spatial velocity of the photon gas measured by \(\mathscr{Z}\).

Contracting \eqref{eq:S eq0}, \eqref{eq:Gibbs-Duhem}, and \eqref{eq:first law} with \(Z_\mu/c^2\), we obtain the Euler relation, the Gibbs-Duhem relation, and the first law as measured by \(\mathscr{Z}\):
\begin{eqnarray*}
  \epsilon_Z = \frac{1}{k_B \gamma \beta} s_Z - P + \frac{1}{c^2} q^{\mu}_Z
  u_{\mu}, & & \text{(Euler relation)}\\
  s_Z \, \mathrm{d} \left( \frac{1}{k_B \beta \gamma} \right) - \mathrm{d} P = 0, & &
  \text{(Gibbs-Duhem relation)}\\
  \mathrm{d} \epsilon_Z = \frac{1}{k_B \beta \gamma} \mathrm{d} s_Z + \frac{1}{c^2}
  u_{\mu} \mathrm{d} q^{\mu}_Z, & & \text{(first law)}
\end{eqnarray*}
where \(\epsilon_Z = c^{-2} Z_{\mu} Z_{\nu} T^{\mu \nu}\) is the energy density,
\(s_Z = - c^{-2} Z_{\mu} S^{\mu}\) is the entropy density, and \(q^{\mu}_Z = - T^{\rho \nu}
Z_{\nu} \Delta^{\mu}_{\enspace \rho}\) is the spatial energy flux measured by \(\mathscr{Z}\). Since the chemical potential of the photon gas is zero (\(\alpha=0\)), the particle number density does not appear. We can identify \((k_B \beta \gamma)^{-1}\) as the temperature measured by \(\mathscr{Z}\), \(T_Z = (k_B \beta \gamma)^{-1}\).

Due to relativistic effects, the Euler relation and the first law measured by a non-comoving observer contain an additional term arising from the motion of the system relative to the observer. The additional term in the first law can be interpreted as the work done to change the momentum density,
\begin{equation}
  \frac{1}{c^2} u_{\mu} \mathrm{d} q_Z^{\mu}
  = \frac{1}{c^2} u_{\mu} \frac{\mathrm{d} q_Z^{\mu}}{\mathrm{d} \tau} \mathrm{d} \tau
  = f^{\mu} u_{\mu} \mathrm{d} \tau,
\end{equation}
where \(q_Z^{\mu}/c^2\) is the spatial momentum density,
\(f^{\mu} = \mathrm{d} q_Z^{\mu}/(c^2 \mathrm{d} \tau)\) is the spatial force density measured by \(\mathscr{Z}\), and \(\tau\) is the proper time of \(\mathscr{Z}\). Thus \(f^{\mu} u_{\mu} \mathrm{d} \tau\) is the work density measured by \(\mathscr{Z}\).

\subsection{Thermal radiation from the equilibrium disk}

Having established that the photon gas and the massive particles share a single
timelike Killing vector field $\mathcal{B}^\mu$ in equilibrium, we would like to see this structure at work
in the Kerr spacetime, of which the line element in Boyer-Lindquist coordinates
is given by
\begin{equation}
  \mathrm{d}s^2 = -\left( 1 - \frac{r_g r}{\rho^2} \right) c^2 \mathrm{d}t^2
  - \frac{2 a r_g r \sin^2 \theta}{\rho^2} c\, \mathrm{d}t\, \mathrm{d}\phi
  + \frac{\Sigma^2}{\rho^2} \sin^2 \theta\, \mathrm{d}\phi^2
  + \rho^2 \left( \frac{\mathrm{d}r^2}{\Delta} + \mathrm{d}\theta^2 \right),
\end{equation}
where $\rho^2 = r^2 + a^2 \cos^2 \theta$, $\Delta = r^2 - r_g r + a^2$,
$\Sigma^2 = (r^2 + a^2)^2 - a^2 \Delta \sin^2 \theta$, and $r_g = 2GM/c^2$,
$a = J/(Mc)$. In Kerr spacetime the timelike Killing vector fields are
generated by $\partial_t$ and $\partial_\phi$, so the equilibrium vector field
takes the form
\begin{equation}
  \mathcal{B}^\mu = \beta_0 \left[ (\partial_t)^\mu + \Omega
  (\partial_\phi)^\mu \right],
\end{equation}
where \(\beta_0\) and \(\Omega\) are constants. The constant \(\Omega\) is the
angular velocity of the coupled system, independent of position, the statement of
rigid rotation. 
The constant \(\beta_0\) sets the overall temperature scale. For a comoving observer,
the local thermodynamic relations derived above hold unchanged. For a set of comoving
observers, one at each point, using \(\mathcal{B}^\mu = \beta U^\mu\) with
\(U^\mu U_\mu = -c^2\), we have \(\beta = |\mathcal{B}|/c\),
where \(|\mathcal{B}| = \sqrt{-\mathcal{B}^\mu\mathcal{B}_\mu}\). Thus all position dependence
of the temperature is carried by the norm $|\mathcal{B}|$, which in Kerr spacetime depends
on both $r$ and $\theta$. For simplicity, we consider a geometrically thin disk in the
equatorial plane $\theta = \pi/2$, so that the temperature varies with $r$ alone. The disk
is also taken to be optically thick, which ensures that the photons exchange energy
frequently with the disk matter and equilibrium is attained. In this case we have
\begin{equation}
  \beta (r) = \beta_0 \sqrt{F (r)},
\end{equation}
where
\begin{equation}
  F (r) \equiv -\frac{\mathcal{B}^\mu \mathcal{B}_\mu}{\beta_0^2 c^2}
  = 1 - \frac{a^2 \Omega^2}{c^2} - \frac{r_g}{r} \left( \frac{a \Omega}{c} -
  1 \right)^2 - \frac{\Omega^2 r^2}{c^2}.
\end{equation}
For \(\Omega \neq 0\), the term \(-\Omega^2 r^2/c^2\) dominates at large \(r\), so \(F(r)\) becomes negative beyond a certain radius. The equilibrium region is then bounded by the light surfaces on which \(\mathcal{B}^\mu\) becomes null, and \(\beta_0\) and \(\Omega\) are fixed by boundary conditions within this region, such as the temperature and angular velocity at the inner edge of the disk. As a consistency check, consider the double limit \(a = 0\) and \(\Omega = 0\), where the profile reduces to \(F = 1 - r_g/r\), so that \(T = T_\infty/\sqrt{1 - r_g/r}\), with \(T_\infty \equiv 1/(k_B \beta_0)\). This is the Tolman law for a static observer in Schwarzschild spacetime.

Inside the disk, the photon gas is coupled to the massive particles, so the temperature
profile is determined. Outside the disk, the photon gas becomes collisionless and is
therefore governed by the Liouville equation alone. The two regimes meet at the disk surface,
so the radiation leaving the disk is fixed by the boundary condition there.
We now analyze the outgoing photon gas at the surface.
To this end, we establish a local orthonormal tetrad frame \((e_{\hat{a}})^\mu\)
such that \(U^\mu = c(e_{\hat{0}})^\mu\) is the fluid proper velocity, and
\begin{equation}
  (e^{\hat{2}})_\mu = (g^{\theta\theta})^{-1/2} (\mathrm{d}\theta)_\mu
  = \rho (\mathrm{d}\theta)_\mu
\end{equation}
defines the unit poloidal frame vector. The tetrad momentum poloidal component
$p^{\hat{2}}$ is related to the coordinate momentum via
$p^{\hat{2}} = p^\mu (e^{\hat{2}})_\mu = \rho p^\theta$. Consequently, the
outgoing condition $p^\theta > 0$ translates to $p^{\hat{2}} > 0$.
Parameterizing the spatial momentum in 3-dimensional spherical coordinates as
\begin{equation}
  p^{\hat{2}} = p^{\hat{0}} \cos \phi_1, \quad
  p^{\hat{1}} = p^{\hat{0}} \sin \phi_1 \cos \phi_2, \quad
  p^{\hat{3}} = p^{\hat{0}} \sin \phi_1 \sin \phi_2,
\end{equation}
the constraint $p^{\hat{2}} > 0$ restricts the polar angles to
$\phi_1 \in (0, \pi/2)$, $\phi_2 \in (0, 2\pi)$.

The invariant momentum volume element takes the form
$\boldsymbol{\varpi}_\gamma = \mathrm{d}^3 p/|p_{\hat{0}}| = p^{\hat{0}}
\mathrm{d}p^{\hat{0}} \mathrm{d}\Omega_2$, where $\mathrm{d}\Omega_2 =
\sin \phi_1 \mathrm{d}\phi_1 \mathrm{d}\phi_2$ is the volume element at the
unit 2-sphere. Performing integration over the outgoing hemisphere
($p^\theta > 0$) yields the standard angular integrals
\begin{align}
  \int_{p^\theta > 0} \mathrm{d}\Omega_2 = 2\pi, \quad~
  \int_{p^\theta > 0} n^{\hat{i}} \mathrm{d}\Omega_2 =
  \pi \delta^{\hat{i}}_{\ \hat{2}}, \quad~
  \int_{p^\theta > 0} n^{\hat{i}} n^{\hat{j}} \mathrm{d}\Omega_2 =
  \frac{2\pi}{3} \delta^{\hat{i}\hat{j}}.
\end{align}
Decomposing the energy-momentum tensor into projections along the fluid rest frame
and the poloidal direction, and using the angular integrals, we obtain the outgoing
energy-momentum tensor in the positive \(\theta\) direction,
\begin{equation}
  T^{\mu\nu} = c \int_{p^\theta > 0} \boldsymbol{\varpi}\, p^\mu p^\nu f_\gamma
  = \frac{\mathfrak{g}_\gamma}{h^3}\frac{2\pi}{c^3\beta^4}\mathcal{I}_3
  \left(\frac{1}{c^2}U^\mu U^\nu + \frac{1}{c}U^{(\mu}(e_{\hat{2}})^{\nu)} + \frac{1}{3}\Delta^{\mu\nu}\right),
  \label{eq:Tmunu of disk}
\end{equation}
where \(\mathcal{I}_3 = \pi^4/15\).

In the local rest frame, the outgoing photon gas is isotropic at the local
temperature $T$. The
energy flow measured by the comoving observer is $\bar{E}^\mu = -T^{\mu\nu}
U_\nu$, and we find that its poloidal component $\bar{E}^{\hat{2}}$ is locally the standard
Stefan-Boltzmann flux
\begin{equation}
  \bar{E}^{\hat{2}} = \sigma T^4, \qquad \sigma = \frac{2\pi^5 k_B^4}{15 h^3
  c^2}.
\end{equation}
This confirms that the equilibrium photon gas reproduces the known blackbody result in the local rest frame.

The energy flow above is measured by an observer comoving with the fluid. To characterize the
radiation in a locally non-rotating frame, and more generally to define integrals of radiation quantities
over a spatial hypersurface, a hypersurface-orthogonal observer is required.
We therefore consider a family of zero-angular-momentum observers
(ZAMO). On the equatorial plane, the proper velocity of ZAMO is
\begin{align}
  Z_\mu = \left(-c \sqrt{\frac{r\Delta}{A}}, 0, 0, 0\right), \qquad A \equiv r^3 + a^2(r_g + r).
\end{align}
Then the proper velocity of the fluid $U^\mu$ can be decomposed into $U^\mu = \gamma (Z^\mu + u^\mu)$ where
\begin{align}
  \gamma = -\frac{1}{c^2} Z_\nu U^\nu = \sqrt{\frac{r\Delta}{A F}}, \qquad
  u^\mu = \left( 0, 0, 0, \frac{\Omega A - a r_g c}{\sqrt{r A \Delta}}
  \right).
\end{align}
The corresponding energy flow measured by the ZAMO is
\begin{equation}
  E^\mu = -T^{\mu\nu} Z_\nu = \frac{\mathfrak{g}_\gamma}{h^3} \frac{2\pi}{c^3 \beta^4}
  \mathcal{I}_3 \left[ \left( \frac{4}{3} \gamma^2 - \frac{1}{3} \right) Z^\mu
  + \frac{4}{3} \gamma^2 u^\mu + \frac{c}{2 r} \gamma
  \delta^\mu_{\ \theta} \right].
\end{equation}

\begin{figure}[!h]
  \centering
  \includegraphics[width=0.97\linewidth]{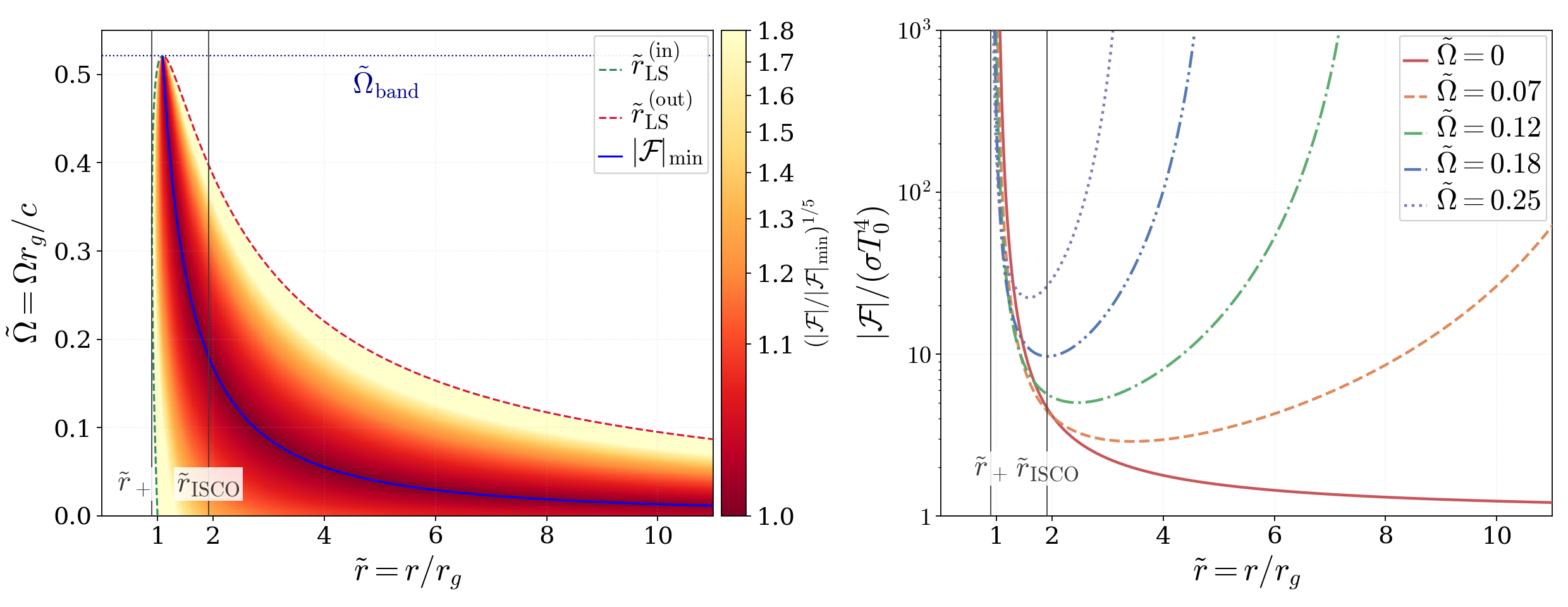}
  \captionsetup{font=footnotesize, labelfont=bf}
  \caption{Magnitude of the spatial energy flux of the equilibrium photon gas on the
  equatorial plane of the Kerr spacetime, for $a / r_g = 0.3$.
  (a) The equilibrium band in the $(r / r_g, \Omega r_g / c)$ plane,
  coloured by $(|\mathcal{F}| / |\mathcal{F}|_{\min}(\Omega))^{1/5}$.
  The blue curve traces the minimum, and the green and red curves are the
  inner and outer light surfaces bounding the band. (b) The absolute flux
  $|\mathcal{F}| / (\sigma T_0^4)$ as a function of
  $r / r_g$ for $\Omega r_g / c = 0, 0.07, 0.12, 0.18, 0.25$.
  Each profile diverges as $F^{-3}$ at the light surfaces,
  and the minimum rises with $\Omega$.}
  \label{fig:flux_mag}
\end{figure}

To interpret this result, we decompose the energy flow measured by the ZAMO
with respect to its proper velocity $Z^{\mu}$,
\begin{equation}
  E^{\mu} = \epsilon_Z Z^{\mu} + \mathcal{F}^{\mu},
  \label{eq:decomp}
\end{equation}
where $\epsilon_Z \equiv - c^{-2} Z_{\mu} E^{\mu}$ is the energy density measured by the ZAMO and
$\mathcal{F}^{\mu} \equiv E^{\mu} - \epsilon_Z Z^{\mu}$ is the spatial energy flux. Substituting the explicit form of $E^{\mu}$ we obtain
\begin{equation}
  \epsilon_Z = \frac{2 \sigma}{3 c} \left( 4 \gamma^2 - 1 \right) T^4, \qquad
  \mathcal{F}^{\mu} = \frac{2 \sigma T^4}{c} \left[ \frac{4}{3} \gamma^2 u^{\mu} + \frac{c}{2 r} \gamma \, \delta^{\mu}_{\ \theta} \right].
  \label{eq:epsZ-Fvec}
\end{equation}
The magnitude of the spatial flux follows directly,
\begin{equation}
  |\mathcal{F}| = \sqrt{\mathcal{F}^{\mu} \mathcal{F}_{\mu}}
  = \sigma T^4 \gamma \, \frac{\sqrt{64 \gamma^2 - 55}}{3},
  \label{eq:absF}
\end{equation}
which is visualized in Fig.~\ref{fig:flux_mag}, where $|\mathcal{F}|$ is measured in units of the fiducial Stefan--Boltzmann flux $\sigma T_0^4$ with $T_0 = 1 / (k_B \beta_0)$. Physically, \(|\mathcal{F}|\) is the energy carried across unit area of a spatial hypersurface per unit proper time of ZAMO, which corresponds to the local cooling rate of the disk.
The equilibrium configuration exists only where $\mathcal{B}^{\mu}$ is timelike, $F (r) > 0$; the band in Fig.~\ref{fig:flux_mag} is therefore bounded by the two light surfaces, where $\mathcal{B}^{\mu}$ becomes null.

\begin{figure}[!h]
  \centering
  \includegraphics[width=0.97\linewidth]{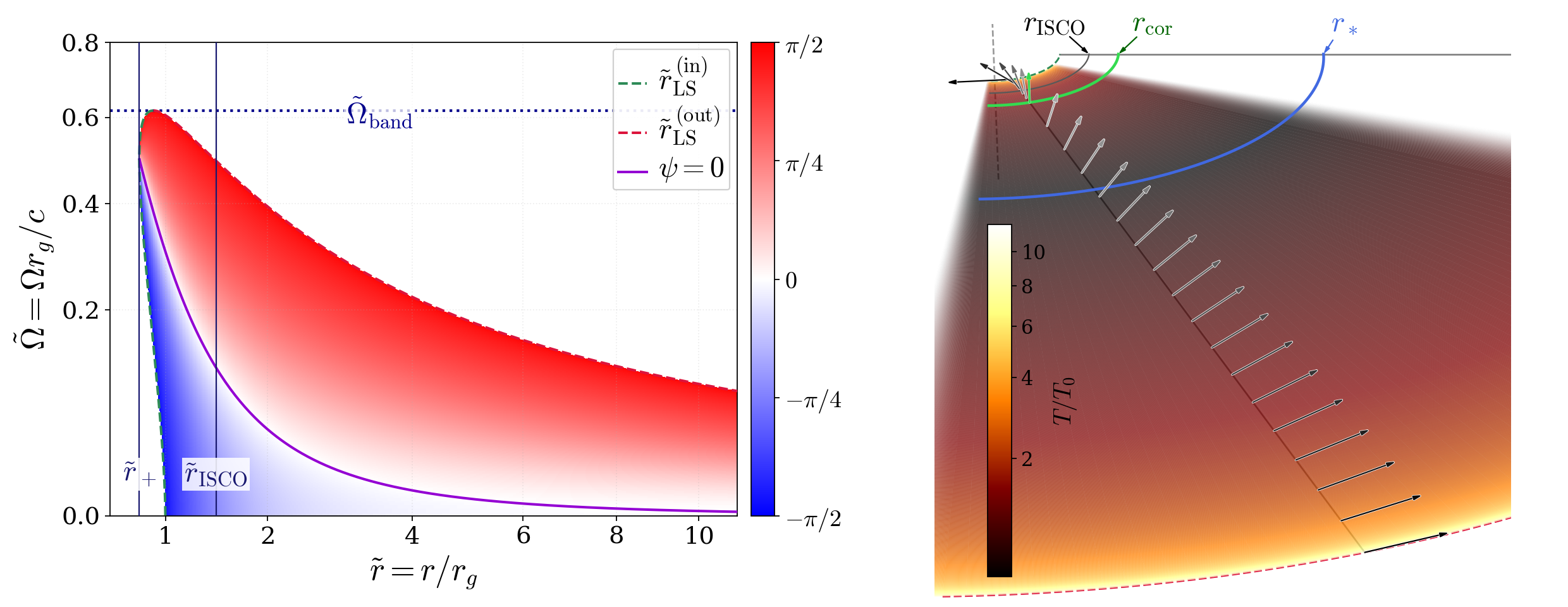}
  \captionsetup{font=footnotesize, labelfont=bf}
  \caption{Direction of the spatial energy flux of the equilibrium photon gas on the
  equatorial plane of the Kerr spacetime, for $a / r_g = 0.4$. (a) The tilt angle
  $\psi$ of the flux away from the disk normal, in the $(r / r_g, \Omega r_g / c)$ plane.
  The colour runs from blue ($\psi = -\pi/2$, the fluid lagging behind the ZAMO)
  through white ($\psi = 0$ on the corotation line $\gamma = 1$) to red
  ($\psi = +\pi/2$, the fluid leading the ZAMO). The band is bounded by the inner
  and outer light surfaces, where $\psi \to \mp \pi/2$.
  (b) A wedge of the disk at $\Omega r_g / c = 0.05$, coloured by the temperature
  $T / T_0 = F^{-1/2}$, which diverges at the light surfaces and is minimal at
  $r_*$. Along a radial line the arrows point along the
  local flux direction, tilting backward inside and forward outside
  the corotation radius, with length equal to the anisotropy factor $\xi$.}
  \label{fig:flux_dir}
\end{figure}

For the direction of the flux, the radial component vanishes, and the poloidal and azimuthal components in the orthonormal frame are given by
\begin{equation}
  \mathcal{F}^{\hat{2}} = \sigma T^4 \gamma, \qquad
  \mathcal{F}^{\hat{3}} = \frac{8}{3} \sigma T^4 \gamma \sqrt{\gamma^2 - 1} \, \operatorname{sgn}(u^{\phi}).
  \label{eq:Fcomp}
\end{equation}
The poloidal component is the Stefan--Boltzmann flux boosted by the Lorentz factor, and the azimuthal component arises from the azimuthal motion of the fluid relative to the ZAMO. The tilt angle $\psi$ away from the disk normal toward the azimuthal direction is therefore
\begin{equation}
  \tan \psi \equiv \frac{\mathcal{F}^{\hat{3}}}{\mathcal{F}^{\hat{2}}}
  = \frac{8}{3} \operatorname{sgn}(u^{\phi}) \sqrt{\gamma^2 - 1}.
  \label{eq:psi}
\end{equation}
The sign is set by the direction of the azimuthal motion of the fluid relative to the ZAMO. Inside the corotation radius $r_{\mathrm{cor}}$, where $\Omega A < a r_g c$, the fluid lags behind the ZAMO and $\psi < 0$. At $r_{\mathrm{cor}}$ the fluid is at rest relative to the ZAMO, $\gamma = 1$, and the flux points along the disk normal, $\psi = 0$. Outside $r_{\mathrm{cor}}$ the fluid leads the ZAMO and $\psi > 0$, with $\psi \to \pm \pi/2$ on the two light surfaces. This direction is visualized in Fig.~\ref{fig:flux_dir}.

A single quantity summarizes how much of the ZAMO energy is carried as directed flux rather than stored as isotropic energy density. We define the anisotropy factor
\begin{equation}
  \xi \equiv \frac{|\mathcal{F}|}{\epsilon_Z c}
  = \frac{\gamma \sqrt{64 \gamma^2 - 55}}{2 \left( 4 \gamma^2 - 1 \right)},
  \label{eq:xi}
\end{equation}
It equals the mean projection of the photon directions on the flux direction.
It is bounded by $1/2 \leqslant \xi \leqslant 1$. The lower bound is reached for an isotropic distribution over the outgoing hemisphere, $\xi = 1/2$, the value that reproduces the Stefan--Boltzmann law in the comoving frame. The upper bound $\xi = 1$ is reached for a perfectly collimated beam, for which the full energy density flows at the speed of light. As the fluid moves faster relative to the ZAMO the aberration beams the radiation forward, and $\xi$ grows from $1/2$ toward $1$. In the 3D panel of Fig.~\ref{fig:flux_dir} this beaming is the arrow length. 

\begin{figure}[!h]
  \centering
  \includegraphics[width=0.7\linewidth]{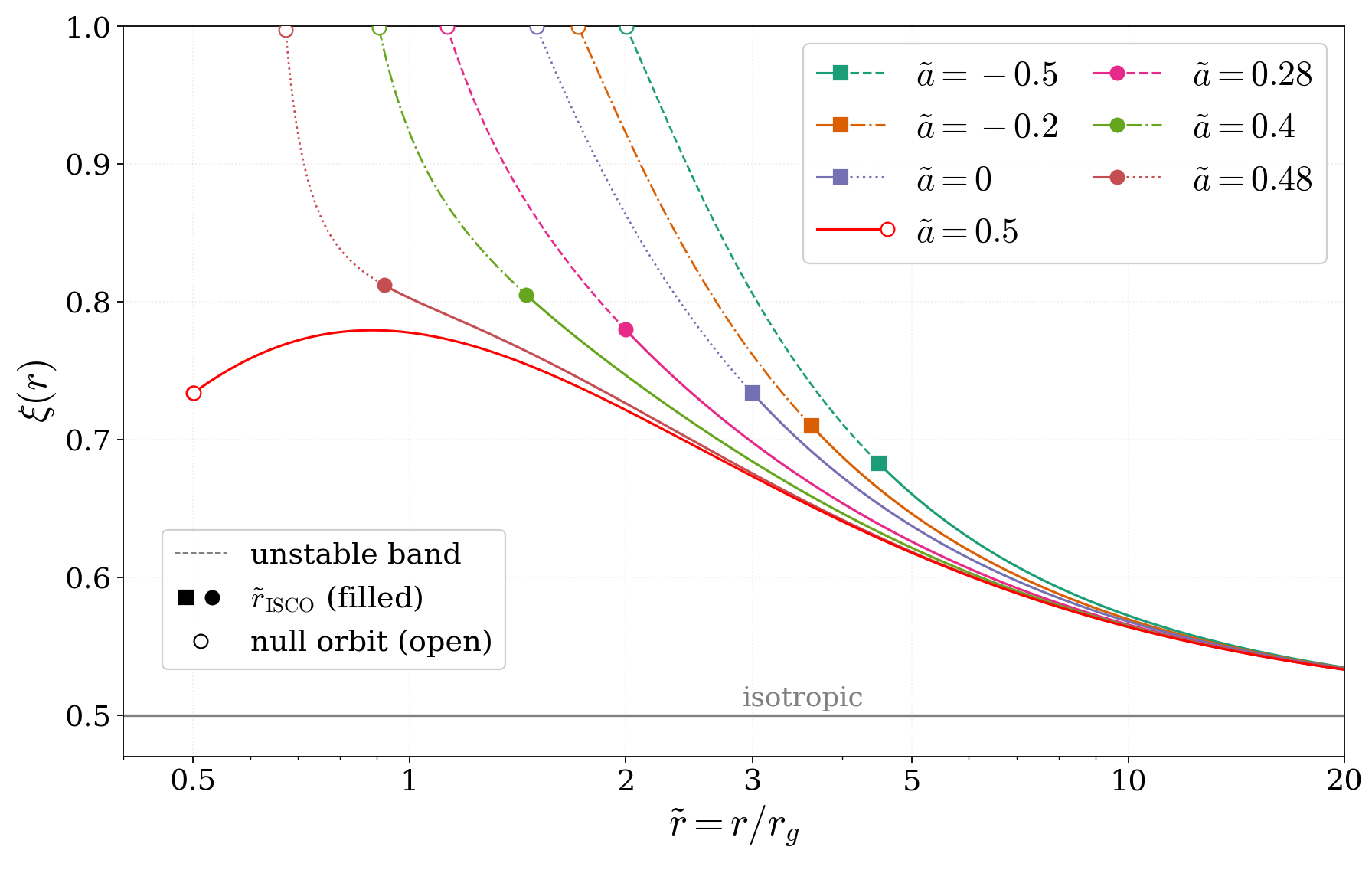}
  \captionsetup{font=footnotesize, labelfont=bf}
  \caption{The anisotropy factor \(\xi(r)\) of the ZAMO energy flux for a local thermodynamic equilibrium Keplerian
  disk on the equatorial plane of the Kerr spacetime, over a family of spins
  \(\tilde{a} = a/r_g = -0.5\), \(-0.2\), \(0\), \(0.28\), \(0.4\), \(0.48\), \(0.5\)
  (negative \(=\) retrograde, positive \(=\) prograde). Each curve runs from the null
  orbit, where $\mathcal B^\mu$ becomes null, through the ISCO to
  \(r = 20\,r_g\). The dashed segment between the null orbit and the ISCO marks
  the unstable band. The retrograde curves have larger inner radii and their
  inner-edge \(\xi\) decreases with \(|a|\), whereas the prograde curves extend
  inward toward the null orbit and their inner-edge \(\xi\) increases with \(a\),
  until at the extremal spin \(a/r_g = 0.5\) the ISCO reaches the null orbit and
  the inner-edge \(\xi\) drops to a finite value.}
  \label{fig:keplerflux}
\end{figure}

The outgoing thermal radiation from the disk is characterized by its energy flow.
For equilibrium disk, this energy flow \eqref{eq:epsZ-Fvec} measured by ZAMO, together
with its magnitude \eqref{eq:absF}, direction \eqref{eq:psi}, and anisotropy factor
\eqref{eq:xi}, complete the analysis at the disk surface. Both the
temperature profile and the energy flux, however, rest on the equilibrium assumption,
which requires $\mathcal{B}^\mu$ to be a Killing vector field and hence forces rigid rotation,
whereas a realistic accretion disk rotates differentially. 
As a first step away from global equilibrium, we relax the global equilibrium assumption
to local thermodynamic equilibrium and consider a Keplerian disk.
In contrast to the freely chosen $\Omega$ of the equilibrium disk, the local angular
velocity is now fixed by the spin $a / r_g$. 
\begin{equation}
  \Omega_K^s (r) = \frac{s \, c \, r_g^{1/2}}{\sqrt{2} \, r^{3/2} + s \, a \, r_g^{1/2}},
  \label{eq:Omegak}
\end{equation}
where \(s = +1\) for a prograde disk and \(s = -1\) for a retrograde one.
The consequence is that $\mathcal{B}^{\mu} = \beta_0 \left[ (\partial_t)^{\mu}
+ \Omega_K^s \, (\partial_{\phi})^{\mu} \right]$ is no longer a Killing vector field,
so the Tolman law no longer holds and the temperature profile is left undetermined.
To local equilibrium order, however, the form of the energy-momentum tensor
\eqref{eq:Tmunu of disk} is unchanged, and the photon gas remains isotropic in local
fluid frame. The anisotropy factor $\xi$ therefore depends on $r$ only through the
boost $\gamma$, and that boost is fixed by the local kinematics, Eq.~\eqref{eq:Omegak}.
Thus $\xi (r)$ remains fully determined by the spin $a / r_g$ alone, even though the
temperature is not. Fig.~\ref{fig:keplerflux} shows $\xi (r)$ for a family of spins
from retrograde $a / r_g = -0.5$ to extremal prograde $a / r_g = 0.5$.

Physically, $\xi$ measures beaming. The photon gas is isotropic in the local fluid frame,
and $\xi$ quantifies how strongly that isotropy is deformed by the orbital motion
as seen by the ZAMO. The distribution function underlying these results, however,
is only the zeroth-order solution and does not by itself satisfy the Boltzmann equation.
Moreover, a real accretion disk is a magnetized plasma of charged particles,
whereas the massive sector considered here has been neutral. Nevertheless,
the covariant framework is established, once the microscopic rates are specified,
one can go beyond the zeroth-order distribution and find the corrections order by order.
To extend the framework to the charged and magnetized case, we present the generalized
collision integrals in the next section.

\section{The generalization of collision integral}\label{sec:The
generalization of collision integral}
In Section \ref{sec:Relativistic kinetic theory for null particles}, we considered only photon absorption-emission and Compton scattering, without reactions between massive particles. Real systems such as accretion disks involve complex electromagnetic processes, including ionization and recombination, and the kinetic theory of null particles has applications beyond photons. We therefore generalize the collision integral to arbitrary multi-species reactions involving both null and massive particles, including charged particles in an electromagnetic field, providing the framework for pair production, ionization equilibrium, and neutrino transport.

Considering the most general collision process
\begin{equation}
  p_{a_1} + \cdots + p_{a_{n_0}} + p_{A_1} + \cdots + p_{A_{n_m}} \to
  p_{b_1} + \cdots + p_{b_{n_0'}} + p_{B_1} + \cdots + p_{B_{n_m'}},
\end{equation}
where $n_0$ null particles and $n_m$ massive particles collide, resulting in
$n_0'$ null particles and $n_m'$ massive particles, with subscripts $a_i, b_j$
denoting different types of null particles, and subscripts $A_i, B_j$
representing different types of massive particles. The corresponding
transition rate $W$, under the constraint of momentum conservation, satisfies
\begin{equation}
  W \left( \sum_i^{n_0} p_{a_i} + \sum_j^{n_m} p_{A_j} \to
  \sum_{i'}^{n_0'} p_{b_{i'}} + \sum_{j'}^{n_m'} p_{B_{j'}} \right) \propto
  \delta^{d + 1} \left( \sum_i^{n_0} p^{\mu}_{a_i} + \sum_j^{n_m}
  p^{\mu}_{A_j} - \sum_{i'}^{n_0'} p^{\mu}_{b_{i'}} - \sum_{j'}^{n_m'}
  p^{\mu}_{B_{j'}} \right) . \label{eq:w delta}
\end{equation}

We introduce the compact wedge-product notation
\begin{equation}
  \bigwedge_i^n \ensuremath{\boldsymbol{\eta}}_i :=
  \ensuremath{\boldsymbol{\eta}}_1 \wedge \ensuremath{\boldsymbol{\eta}}_2
  \wedge \cdots \wedge \ensuremath{\boldsymbol{\eta}}_n .
\end{equation}
Through a construction process entirely similar to the previous section, we can
derive the Boltzmann equations for null particles and massive particles. The
Boltzmann equation for null particles $a_k$ is
\begin{equation}
  L_{a_k} [f_{a_k}] = p_{a_k}^{\mu}  \cfrac{\partial f_{a_k}}{\partial
  x^{\mu}} - \Gamma^{\mu}_{\alpha \beta} p_{a_k}^{\alpha} p_{a_k}^{\beta} 
  \cfrac{\partial f_{a_k} }{\partial p_{a_k}^{\mu}} = \underset{\text{involve
  $a_k$}}{\underset{\text{collision}}{\sum}} \mathcal{C}_\mathfrak{i} (x, p_{a_k}),
  \label{eq:generalized Boltzmann equation for null particle}
\end{equation}
where the summation is over the collision processes involving the null
particle $a_k$, and the collision integral is
\[ \mathcal{C}_\mathfrak{i} (x, p_{a_k}) = \int \bigwedge^{n_0}_{i \neq k} \boldsymbol{\varpi}_{a_i} \wedge \bigwedge_j^{n_m} \boldsymbol{\varpi}_{A_j} \wedge \bigwedge_{i'}^{n_0'} \boldsymbol{\varpi}_{b_{i'}} \wedge \bigwedge_{j'}^{n_m'} \boldsymbol{\varpi}_{B_{j'}} \; [\mathcal{W}_{(\text{out} \to \text{in})} - \mathcal{W}_{(\text{in} \to \text{out})}] , \]
where $\text{in} = \{a_i\} \cup \{A_j\}$, $\text{out} = \{b_{i'}\} \cup \{B_{j'}\}$, and the subscript $\mathfrak{i}$ in $\mathcal{C}_\mathfrak{i} (x, p_{a_k})$ is used to distinguish
different collision processes.

The Boltzmann equation for massive particles $A_l$ is
\begin{equation}
  L_{A_l} [f_{A_l}] = p_{A_l}^{\mu}  \cfrac{\partial f_{A_l}}{\partial
  x^{\mu}} + \left( q_{A_l} F^{\mu}_{\enspace \nu} p_{A_l}^{\nu} -
  \Gamma^{\mu}_{\alpha \beta} p_{A_l}^{\alpha} p_{A_l}^{\beta} \right) 
  \cfrac{\partial f_{A_l} }{\partial p_{A_l}^{\mu}} = \underset{\text{involve
  $A_l$}}{\underset{\text{collision}}{\sum}} \mathcal{C}_\mathfrak{i} (x, p_{A_l}),
  \label{eq:generalized Boltzmann equation for massive particle}
\end{equation}
where the summation is over the collision processes involving the massive
particle $A_l$, and the collision integral is
\[ \mathcal{C}_\mathfrak{i} (x, p_{A_l}) = \int \bigwedge^{n_0}_{i} \boldsymbol{\varpi}_{a_i} \wedge \bigwedge_{j \neq l}^{n_m} \boldsymbol{\varpi}_{A_j} \wedge \bigwedge_{i'}^{n_0'} \boldsymbol{\varpi}_{b_{i'}} \wedge \bigwedge_{j'}^{n_m'} \boldsymbol{\varpi}_{B_{j'}} \; [\mathcal{W}_{(\text{out} \to \text{in})} - \mathcal{W}_{(\text{in} \to \text{out})}] . \]

Similarly, using equations \eqref{eq:divT} and \eqref{eq:w delta}, we can
demonstrate that under the evolution governed by the Boltzmann equations
\eqref{eq:generalized Boltzmann equation for null particle} and
\eqref{eq:generalized Boltzmann equation for massive particle}, the divergence
of the total energy-momentum tensor (including null particles, massive
particles, and the electromagnetic field) is zero. Likewise, for particles
which remain unchanged before and after collisions, the corresponding
divergence of particle flow is also zero.

Through an analysis analogous to that in Section \ref{sec:H-theorem}, we can
demonstrate that for the H-theorem to hold universally, the transition
rates for the forward and reverse processes should satisfy
\begin{equation}
  \frac{W \left( \sum_i^{n_0} p_{a_i} + \sum_j^{n_m} p_{A_j} \to
  \sum_{i'}^{n_0'} p_{b_{i'}} + \sum_{j'}^{n_m'} p_{B_{j'}} \right)}{W \left(
  \sum_{i'}^{n_0'} p_{b_{i'}} + \sum_{j'}^{n_m'} p_{B_{j'}} \to
  \sum_i^{n_0} p_{a_i} + \sum_j^{n_m} p_{A_j} \right)} = \frac{(h^d)^{n_m -
  n_m' + n_0 - n_0'} }{(\mathfrak{g}^{n_0}_{a_k}
  /\mathfrak{g}^{n_0'}_{b_{k'}})  (\mathfrak{g}^{n_m}_{A_l}
  /\mathfrak{g}^{n_m'}_{B_{l'}})} . \label{eq:generic detailed balance W}
\end{equation}
When the entropy production rate of the system vanishes, the system reaches
global equilibrium. In this case, the vanishing of the divergence of the entropy flow can be read as the statement that these combinations are invariant across collisions
\begin{equation}
  \sum_k^{n_0} \log \frac{h^d f_{a_k}}{\mathfrak{g}_{a_k} f_{a_k}^{\ast}} +
  \sum_l^{n_m} \log \frac{h^d f_{A_l}}{\mathfrak{g}_{A_l} f_{A_l}^{\ast}} =
  \sum_{k'}^{n_0'} \log \frac{h^d f_{b_{k'}}}{\mathfrak{g}_{b_{k'}}
  f_{b_{k'}}^{\ast}} + \sum_{l'}^{n_m'} \log \frac{h^d
  f_{B_{l'}}}{\mathfrak{g}_{B_{l'}} f_{B_{l'}}^{\ast}} . \label{eq:Collision
  invariant}
\end{equation}

If the system cannot be decomposed into several subsystems such that every
particle in any subsystem does not interact with particles in other
subsystems (that is, all particles will interact with each other either directly
or indirectly), all components of the system have the same detailed balance state.
Recall that equation \eqref{eq:Collision invariant} represents the conserved
quantities before and after collisions. Therefore, for massive particles,
$\log \left[ (h^d f_{A_l}) / (\mathfrak{g}_{A_l} f_{A_l}^{\ast}) \right]$ can be written as a
linear combination of momentum and scalar fields. For massless particles,
however, there is no corresponding particle number conservation condition, so
$\log \left[ (h^d f_{a_k}) / (\mathfrak{g}_{a_k} f_{a_k}^{\ast}) \right]$ can only be written
as a linear combination of momentum
\[ \log \frac{h^d f_{a_k}}{\mathfrak{g}_{a_k} f_{a_k}^{\ast}} =
   \mathcal{B}_{\mu} p_{a_k}^{\mu}, \quad \log \frac{h^d
   f_{A_l}}{\mathfrak{g}_{A_l} f_{A_l}^{\ast}} = 
   \mathcal{B}_{\mu} p_{A_l}^{\mu} - \alpha_{A_l} . \]
Then the detailed balance distribution function of this system is
\begin{equation}
  f_{a_k} = \frac{\mathfrak{g}_{a_k}}{h^d}
  \frac{1}{\mathrm{e}^{-\mathcal{B}_{\mu} p_{a_k}^{\mu}} - \varsigma_{a_k}},
  \quad f_{A_l} = \frac{\mathfrak{g}_{A_l}}{h^d}
  \frac{1}{\mathrm{e}^{\alpha_{A_l} -\mathcal{B}_{\mu} p_{A_l}^{\mu}} -
  \varsigma_{A_l}}, \label{eq:generic detailed balance f}
\end{equation}
in which the different components of the system have the same
$\mathcal{B}_{\mu}$, i.e., the same temperature and flow velocity, and the
detailed balance condition is
\begin{equation}
  \nabla_{\mu} \alpha_A - q_A F^{\nu}_{\enspace \mu} \mathcal{B}_{\nu} = 0,
  \hspace{1em} \quad \nabla_{(\mu} \mathcal{B}_{\nu)} = 0. \label{eq:generic
  detailed balance condition}
\end{equation}
Additionally, the electromagnetic field $F_{\mu \nu}$ must be stationary
\begin{equation}
  \mathcal{L}_{\mathcal{B}}  \ensuremath{\boldsymbol{F}} = (\mathrm{d}
  \iota_{\mathcal{B}} + \iota_{\mathcal{B}} \mathrm{d})
  \ensuremath{\boldsymbol{F}} = \mathrm{d} \iota_{\mathcal{B}}
  \ensuremath{\boldsymbol{F}} = \frac{1}{q_A} {\mathrm{d} \mathrm{d} \alpha_A}
  = 0, \label{eq:generic detailed balance condition F}
\end{equation}
where $\ensuremath{\boldsymbol{F}} = F_{\mu \nu} \mathrm{d} x^{\mu} \otimes
\mathrm{d} x^{\nu}$, we utilized the Cartan formula for Lie derivative, the
properties of the electromagnetic tensor $\mathrm{d}
\ensuremath{\boldsymbol{F}} = 0$, and the detailed balance condition
\eqref{eq:generic detailed balance condition} rewritten in a differential form
$\mathrm{d} \alpha_A - q_A \iota_{\mathcal{B}} \ensuremath{\boldsymbol{F}} =
0$. For each collision process, the parameter $\alpha$ before and after
the collision must satisfy
\begin{equation}
  \sum_j^{n_m} \alpha_{A_j} = \sum_{j'}^{n_m'} \alpha_{B_{j'}},
  \label{eq:generic detailed balance condition alpha}
\end{equation}
which is known as the chemical equilibrium condition. 

If the system can be decomposed into several subsystems such that every
particle in any subsystem does not interact with particles in other
subsystems, then each subsystem has a different detailed balance state. The
detailed balance distribution function for each component is the same as
equation \eqref{eq:generic detailed balance f} which satisfies corresponding
detailed balance conditions \eqref{eq:generic detailed balance condition},
\eqref{eq:generic detailed balance condition F} and \eqref{eq:generic detailed
balance condition alpha}, but the equilibrium temperatures and flow velocities
of different subsystems may differ. Roughly speaking, if the interaction
between two components of a system is weak enough, it can be approximated that
each component reaches a different equilibrium state within a certain
timescale. For example, in some accretion disk theory, it is commonly assumed that
electrons and ions have different temperatures.
\section{Conclusion}
In this paper, we complete the covariant Boltzmann equation for photons coupled to
massive particles in general curved spacetimes by deriving the collision integrals for
photon-matter interactions.

This equation includes collision integrals for emission-absorption, Compton scattering,
and massive-particle scattering, for which we proved the corresponding conservation laws.
A notable feature of the absorption-emission process is that it changes the photon number,
so the H-theorem does not follow automatically as in number-conserving scattering.
Requiring it to hold determines the relation
\(W_{\text{abs}} = (h^d/\mathfrak{g}_\gamma)W_{\text{emi}}\) between the absorption and
emission transition rates. In the non-relativistic limit this relation reduces to the
standard Einstein relation among the \(A\) and \(B\) coefficients, and it emerges here
as a consequence of the second law of thermodynamics rather than as a separate microscopic
input. The collision integral was further generalized to arbitrary multi-species reactions
involving both null and massive particles, including charged particles in an electromagnetic field.

The covariant thermodynamic relations of the photon gas, together with their
observer-dependent forms, follow from the local equilibrium distribution function.
Detailed balance further requires a shared timelike Killing vector field
\(\mathcal{B}^\mu\), which means that the photon gas admits a timelike proper velocity,
even though individual photons travel along null worldlines. For a comoving observer,
the equilibrium photon gas is precisely the familiar blackbody radiation.
As an illustration, we applied these relations to the thermal radiation from an
optically thick disk in Kerr spacetime, where they reproduce the Tolman law and
the Stefan--Boltzmann flux in the appropriate limits. In local thermodynamic equilibrium
for a Keplerian disk, the anisotropy factor of the emitted flow is fixed by the
black-hole spin alone.
These examples, however, end at the disk surface, beyond which the photon gas
is collisionless and evolves according to the Liouville equation alone.
Such a collisionless evolution is a separate problem. In Kerr spacetime, the geodesic flow
is Liouville integrable, so the cotangent-bundle method~\cite{Rioseco:2016jwc,Mach:2021zqe,
Cai:2022fdu,Liao:2022lza,Shapiro:2023gpe,Li:2023qtd,Liu:2026qic},
via the Hamilton–Jacobi separation, yields the general solution of the
Liouville equation as an arbitrary function of the constants of motion.

These results are a step toward a realistic non-equilibrium
description of accreting systems. The transition rates remain unspecified at
the microscopic level, and a fully non-equilibrium distribution remains out of reach.
The framework nonetheless provides a first-principles foundation for studying radiative
processes in strong gravitational fields, with potential applications to accretion disks,
jets, and other extreme astrophysical environments. 
Natural next steps include specifying
the transition rates from microscopic physics, extending the analysis to magnetized plasmas,
and solving the Boltzmann equation for a non-equilibrium distribution function.

\section*{Acknowledgment}
This work is supported by the National Natural Science Foundation of China (Grant Nos. 12275216, 12247103).

\normalem
\bibliographystyle{utphys}
\bibliography{ref}

\begin{appendices}
  \section{Quantum statistical factors}\label{sec:quantum_statistical_factors}

For a degenerate quantum system, we need to make quantum statistical corrections to the collision integral.
Consider a degenerate quantum system where the occupation number of a certain state is $\mathfrak{n}$. 
For a Bose system, when the creation operator acts on this state,
\[ a^{\dagger} | \mathfrak{n}\rangle = \sqrt{1 +\mathfrak{n}} | \mathfrak{n}+ 1 \rangle.
\]
This indicates that the probability of a Boson system in state $| \mathfrak{n}\rangle$ entering the final state $| \mathfrak{n}+1\rangle$ is proportional to $1 + \mathfrak{n}$.
For a Fermi system, when the creation operator acts on this state,
\[ a^{\dagger} | \mathfrak{n}\rangle = \sqrt{1 -\mathfrak{n}} | \mathfrak{n}+ 1 \rangle.
\]
This means that the probability of a Fermi system in state $| \mathfrak{n}\rangle$ entering the final state $| \mathfrak{n}+1\rangle$ is proportional to $1 -  \mathfrak{n}$. 
The above discussion implies that it is not sufficient to measure the ability to transition from the initial state to the final state only by the transition rate independent of the distribution function for bosons and fermions. We also need to take into account the influence of the occupation number of the final state.

In the following we consider the above discussion within relativistic kinetic theory. 
First, we establish an observer orthogonal frame, such that the proper velocity of the hypersurface-orthogonal observer $Z_{\mu} = - c (e^{\hat{0}})_{\mu}$. Then, the number of particles within a certain spatial region $\Sigma$ is 
\[ \int_{\Sigma} \ensuremath{\boldsymbol{\eta}}_{\Sigma} \left( -
\frac{1}{c^2} Z_{\mu} N^{\mu} \right) = - \frac{1}{c} Z_{\mu} \int_{\Sigma}
\ensuremath{\boldsymbol{\eta}}_{\Sigma} \int_{P_m}
\ensuremath{\boldsymbol{\varpi}} f p^{\mu} = \int_{\Sigma}
\ensuremath{\boldsymbol{\eta}}_{\Sigma} \int_{P_m} \frac{\mathrm{d}^d
p}{p^{\hat{0}}} f p^{\hat{0}} = \int_{\Sigma}
\ensuremath{\boldsymbol{\eta}}_{\Sigma} \int_{P_m} \mathrm{d}^d p f , \]
where $\ensuremath{\boldsymbol{\eta}}_{\Sigma}$ is the volume element of the spatial region $\Sigma$.
This provides another way to interpret the distribution function, that is, the number of particles within the unit phase volume element $\ensuremath{\boldsymbol{\eta}}_{\Sigma} \wedge \mathrm{d}^d p$ is $f$.
Taking into account the degeneracy $\mathfrak{g}$, since the number of states that a unit phase volume element can accommodate is $\mathfrak{g}/h^{d}$, the average occupation number of a single state is $\mathfrak{n} = \mathfrak{g}^{-1} h^d f$.

After the collision process occurs, the probability of the scattering process is affected by the occupation number of the final state. By unifying the descriptions of Bose systems, Fermi systems and Boltzmann systems, the intensity of transition needs to be multiplied by the corresponding quantum correction factor
$f^{\ast} = 1 + \varsigma \mathfrak{n} = 1 + \varsigma \mathfrak{g}^{- 1} h^d f$, where $\varsigma = 0, 1, - 1$ to represent the non-degenerate case, Bose-Einstein statistics, and Fermi-Dirac statistics respectively. Therefore, taking into account the corrections in quantum statistics, the collision integral will be modified to
\begin{equation}
  \mathcal{C} (x, p_1) = \int_{P_m \times P_m \times P_m} \ensuremath{\boldsymbol{\varpi}}_2 
  \ensuremath{\boldsymbol{\varpi}}_3  \ensuremath{\boldsymbol{\varpi}}_4  [W
  (p_3 + p_4 \to p_1 + p_2) f_3 f_4 f_1^{\ast} f_2^{\ast} - W (p_1 + p_2
  \to p_3 + p_4) f_1 f_2 f_3^{\ast} f_4^{\ast}],
\end{equation}

  \section{Derivation of equation \eqref{eq:DN}}\label{sec:Derivation
DN}

For the convenience of expression, the calculation is limited to null
particles. The methodology for massive particles is entirely comparable. To
make the derivation explicit, we briefly introduce some conclusions regarding
the geometry of the tangent bundle here. The Sasaki
metric is a natural metric for the tangent bundle $T M$, which is given by
\begin{equation}
  \hat{g} = g_{\mu \nu} \mathrm{d} x^{\mu} \otimes \mathrm{d} x^{\nu} + g_{\mu
  \nu} \theta^{\mu} \otimes \theta^{\nu},
\end{equation}
where $g_{\mu \nu}$ is the metric on spacetime $M$, and $\theta^{\mu} =
\mathrm{d} p^{\mu} + \Gamma^{\mu}_{\nu \sigma} p^{\nu} \mathrm{d} x^{\sigma}$
is the basis of the covectors in the vertical subspace of the cotangent space
of the tangent bundle. Correspondingly, the tangent space of the tangent bundle
has the dual basis,
\begin{equation}
  e_{\mu} = \frac{\partial}{\partial x^{\mu}} - \Gamma^{\nu}_{\mu \sigma}
  p^{\sigma} \frac{\partial}{\partial p^{\nu}}, \quad \frac{\partial}{\partial
  p^{\mu}},
\end{equation}
that satisfies the orthonormal condition
\begin{equation}
  \mathrm{d} x^{\mu} [e_{\nu}] = \delta^{\mu}_{\enspace \nu}, \quad
  \theta^{\mu} \left[ \frac{\partial}{\partial p^{\nu}} \right] =
  \delta^{\mu}_{\enspace \nu}, \quad \mathrm{d} x^{\mu} \left[
  \frac{\partial}{\partial p^{\nu}} \right] = 0, \quad \theta^{\mu} [e_{\nu}]
  = 0.
\end{equation}
Then, the Liouville vector field can be expressed as
\begin{equation}
  L = p^{\mu} e_{\mu} .
\end{equation}
For a detailed geometric structure of the tangent space of the tangent bundle,
one may refer to chapter 2 of {\cite{Sarbach2013}}.

In previous work \cite{cai2025geometry}, we show that by using the flow
$\mathcal{J}= f L$ on the mass shell $\Gamma_m$, the density of particle
trajectories on mass shell can be expressed as $- \hat{g} (\mathcal{J},
\mathcal{Z}) / c^2$, where $\mathcal{Z}= Z^{\mu} e_{\mu}$ is the normal vector
with the normalization condition $\hat{g} (\mathcal{Z}, \mathcal{Z}) = - c^2$,
orthogonal to the hypersurface $\hat{\Sigma}_t$ in the mass shell
$\Gamma_m$, and $Z^{\mu}$ is orthogonal to the $d$-dimensional spacelike
hypersurface $\Sigma_t$ at the given time $t$ in spacetime, representing the
proper velocity of an observer on $M$. To avoid a complex discussion, here we
will only verify that the density of trajectories $- \hat{g} (\mathcal{J},
\mathcal{Z}) / c^2$ on $\hat{\Sigma}_t$ is compatible with equation
\eqref{eq:N1}. Noting that the relation between $\hat{\Sigma}_t$ and
$\Sigma_t$ is
\[ \hat{\Sigma}_t = \{ (x, p) |x \in \Sigma_t, p \in P_0 \} . \]
Integrating the number density of trajectories over the fiber space $P_0$
then yields the particle number density on $\Sigma_t$
\begin{align*}
  - \frac{1}{c^2} \int \ensuremath{\boldsymbol{\eta}}_{P_0}  \hat{g}
  (\mathcal{J}, \mathcal{Z})  = & - \frac{1}{c^2} c \int
  \ensuremath{\boldsymbol{\varpi}} f \hat{g} (L, \mathcal{Z})\\
   = & - \frac{1}{c^2} Z_{\mu} \left( c \int
  \ensuremath{\boldsymbol{\varpi}} f p^{\mu} \right)\\
   = & - \frac{1}{c^2} Z_{\mu} N^{\mu} .
\end{align*}
In the final step, we substituted the Eq.~\eqref{eq:N1}. $- Z_{\mu}
N^{\mu} / c^2$ is precisely the particle number density on $\Sigma_t$, which
means that we can calculate the change in particle number using the flow
$\mathcal{J}= f L$. This is very helpful for the subsequent proof.

Now we present two lemmas without proof, the proof of which can be found in
standard differential geometry textbooks.

\newtheorem{lemma}{Lemma}
\begin{lemma}
  \label{lemma:1}Given a pseudo-Riemannian manifold $M$ with metric $\hat{g}$,
  the integral of the vector field $\mathcal{J}$ over its codimension-one spacelike
  hypersurface $\hat{\Sigma}$ can be expressed as
  \begin{equation}
    \int_{\hat{\Sigma}} \hat{g} (\mathcal{J}, n) 
    \ensuremath{\boldsymbol{\eta}}_{\hat{\Sigma}} = \int_{\hat{\Sigma}}
    \mathcal{J}^{\mu} n_{\mu}  \ensuremath{\boldsymbol{\eta}}_{\hat{\Sigma}} =
    - \int_{\hat{\Sigma}} \iota_{\mathcal{J}}  \ensuremath{\boldsymbol{\eta}},
  \end{equation}
  where $n$ is the timelike unit normal vector of $\hat{\Sigma}$ satisfying
  $n^{\mu} n_{\mu} = - 1$, $\iota$ denotes the interior product,
  $\ensuremath{\boldsymbol{\eta}}$ is the volume element of $M$, and
  $\ensuremath{\boldsymbol{\eta}}_{\hat{\Sigma}} = \iota_n
  \ensuremath{\boldsymbol{\eta}}$ is the volume element of $\hat{\Sigma}$.
\end{lemma}

\begin{lemma}
  \label{lemma:2}Given a full-rank form $\ensuremath{\boldsymbol{\eta}}$ and a
  vector field $\mathcal{J}= f L$ on the manifold, where $f$ is a scalar field
  and $L$ is a vector field, the Lie derivative of $\ensuremath{\boldsymbol{\eta}}$
  along the vector field $\mathcal{J}^{\mu}$ can be expanded as
  \begin{equation}
    \mathcal{L}_{\mathcal{J}} \ensuremath{\boldsymbol{\eta}} =\mathcal{L}_{f
    L} \ensuremath{\boldsymbol{\eta}} = L [f] \ensuremath{\boldsymbol{\eta}} +
    f\mathcal{L}_L \ensuremath{\boldsymbol{\eta}} .
  \end{equation}
\end{lemma}

Using these two lemmas, we can directly calculate the change in the number of
particle trajectories.
\begin{align}
  \Delta N  = & - \frac{1}{c^2} \int_{\hat{\Sigma}_2 \bigcap \partial
  \hat{V}} \hat{g} (\mathcal{J}, \mathcal{Z})  \ensuremath{\boldsymbol{\eta
  }}_{\hat{\Sigma}} + \frac{1}{c^2} \int_{\hat{\Sigma}_1 \bigcap \partial
  \hat{V}} \hat{g} (\mathcal{J}, \mathcal{Z})  \ensuremath{\boldsymbol{\eta
  }}_{\hat{\Sigma}} \nonumber\\
   = &~ \frac{1}{c} \int_{\hat{\Sigma}_2 \bigcap \partial \hat{V}}
  \iota_{\mathcal{J}}  \ensuremath{\boldsymbol{\eta }}_{\Gamma_0 } -
  \frac{1}{c} \int_{\hat{\Sigma}_1 \bigcap \partial \hat{V}} \iota_{\mathcal{J}} 
  \ensuremath{\boldsymbol{\eta }}_{\Gamma_0 } \nonumber\\
   = &~ \frac{1}{c} \int_{\partial \hat{V}} \iota_{\mathcal{J}} 
   \ensuremath{\boldsymbol{\eta }}_{\Gamma_0 }
  \nonumber\\
   = &~ \frac{1}{c} \int_{\hat{V}} \mathcal{L}_{\mathcal{J}} 
  \ensuremath{\boldsymbol{\eta }}_{\Gamma_0 } \nonumber\\
   = &~ \frac{1}{c} \int_{\hat{V}} \ensuremath{\boldsymbol{\eta }}_{\Gamma_0
  } (L [f] + f\mathcal{L}_L \ensuremath{\boldsymbol{\eta }}_{\Gamma_0 })
  \nonumber\\
   = &~ \int_{\hat{V}} \ensuremath{\boldsymbol{\eta_M}} \wedge
  \ensuremath{\boldsymbol{\varpi}} L [f] , 
\end{align}
where the second step employs Lemma \ref{lemma:1}. The third step relies on
the fact that the integral of the vector field $\mathcal{J}^{\mu} = f L^{\mu}$
along the lateral boundary vanishes (since $\hat{V}$ is bounded by the
integral curves of $L^{\mu}$). The fourth step involves Stokes theorem and
Cartan formula $\mathcal{L}_{\mathcal{J}} \ensuremath{\boldsymbol{\eta}} =
\mathrm{d} \iota_{\mathcal{J}} \ensuremath{\boldsymbol{\eta}} +
\iota_{\mathcal{J}} \mathrm{d} \ensuremath{\boldsymbol{\eta}}$. Progressing to
the fifth step, we utilize Lemma \ref{lemma:2}. The sixth step uses the Liouville
theorem $\mathcal{L}_L \ensuremath{\boldsymbol{\eta }}_{\Gamma_0 } = 0$ and
$\ensuremath{\boldsymbol{\eta }}_{\Gamma_0 } =
\ensuremath{\boldsymbol{\eta}}_M \wedge \ensuremath{\boldsymbol{\eta}}_{P_0} =
c \ensuremath{\boldsymbol{\eta}}_M \wedge \ensuremath{\boldsymbol{\varpi}}$.
  \section{Einstein coefficients}\label{sec:einstein_coefficients}

This appendix presents the explicit connection between the microscopic
transition rates and the traditional phenomenological Einstein $A$ and $B$
coefficients, extended to $d$ spatial dimensions.

Consider a two-level system with a lower energy state $A'$ (degeneracy
$\mathfrak{g}_A'$) and an upper energy state $A$ (degeneracy
$\mathfrak{g}_A$) for massive particles in flat spacetime. In traditional
radiative transfer theory, the temporal change of the upper state particle
number density $n_A$ is written as
\begin{equation}
  \frac{\mathrm{d} n_A}{\mathrm{d} t} = - A_{A \to A'} n_A - B_{A \to A'} \rho
  (\nu) n_A + B_{A' \to A} \rho (\nu) n_A', \label{eq:app_traditional_rate}
\end{equation}
where $\rho (\nu)$ is the spectral energy density of the radiation field at
frequency $\nu$, $A_{A \to A'}$ denotes the spontaneous emission coefficient, $B_{A \to A'}$ the stimulated emission coefficient, and $B_{A' \to A}$ the stimulated absorption coefficient.

The proper velocity of the particle system is denoted as $U^{\mu}$. We
consider a system in thermal equilibrium in the low velocity regime, where the
massive particles $A$ and $A'$ move at non-relativistic speeds relative to the
system ($p_A^{\hat{0}} = - \dfrac{1}{c} U_{\mu} p_A^{\mu} = m_A c +\mathcal{O}
(v_A^2)$).

We assume that the massive species $A$ and $A'$ follow the classical
Maxwell-Boltzmann distribution, while the photons follow the Bose-Einstein
distribution:
\begin{equation}
  f_A = \frac{\mathfrak{g}_A}{h^d} \mathrm{e}^{- \alpha - \beta E_A}, \quad
  f_A' = \frac{\mathfrak{g}_A'}{h^d} \mathrm{e}^{- \alpha - \beta E_A'}, \quad
  f_{\gamma} = \frac{\mathfrak{g}_{\gamma}}{h^d}  \frac{1}{\mathrm{e}^{\beta h
  \nu} - 1} . \label{eq:app_distributions}
\end{equation}
where $E_A = cp_A^{\hat{0}} = - U_{\mu} p_A^{\mu}$, $h \nu =
cp_{\gamma}^{\hat{0}} = - U_{\mu} p_{\gamma}^{\mu}$ and $\mathfrak{g}_{\gamma}$ is the degeneracy of photons.

Assuming that only the processes of particles absorbing and emitting photons
exist in the system, the corresponding Boltzmann equation for upper energy
state $A$ is
\begin{equation}
  L [f_A] =\mathcal{C}_{\mathrm{rad}} (x, p_A),
\end{equation}
where the collision integral is
\begin{eqnarray}
  \mathcal{C}_{\mathrm{rad}} (x, p_A) & = & \int
  \ensuremath{\boldsymbol{\varpi}}_{\gamma}
  \ensuremath{\boldsymbol{\varpi}}_A'  [ W_{\mathrm{abs}} (p_{\gamma} + p_A'
    \to p_A) f_{\gamma} f_A' - W_{\mathrm{emi}} (p_A \to p_A' +
  p_{\gamma}) f_A f_{\gamma}^{\ast}] . 
\end{eqnarray}
According to Eq.~\eqref{eq:generic detailed balance W}, the microscopic
transition rates for emission $W_{\mathrm{emi}}$ and absorption
$W_{\mathrm{abs}}$ satisfy the relation
\begin{equation}
  W_{\mathrm{abs}}  (p_{\gamma} + p_A' \to p_A) =
  \frac{\mathfrak{g}_A}{\mathfrak{g}_A'}  \frac{h^d}{\mathfrak{g}_{\gamma}}
  W_{\mathrm{emi}}  (p_A \to p_A' + p_{\gamma}) . \label{eq:app_micro_db}
\end{equation}

The energy-momentum tensor of photons is given by
\begin{equation}
  T^{\mu \nu} = c \int \ensuremath{\boldsymbol{\varpi}}_{\gamma}
  p_{\gamma}^{\mu} p_{\gamma}^{\nu} f_{\gamma} = \frac{\mathcal{A}_{d -
  1}}{c^d}  \int \mathrm{d} \nu \frac{\mathfrak{g}_{\gamma}}{h^d}  \frac{h^{d
  + 1} \nu^d}{\mathrm{e}^{\beta h \nu} - 1}  \left( \frac{1}{c^2} U^{\mu}
  U^{\nu} + \frac{1}{d} \Delta^{\mu \nu} \right),
\end{equation}
where $\ensuremath{\boldsymbol{\varpi}}_{\gamma} = (p_{\gamma}^{\hat{0}})^{d -
2} \mathrm{d} p_{\gamma}^{\hat{0}} \mathrm{d} \Omega_{d - 1}$ is the
Lorentz-invariant volume element for photons, and $\mathcal{A}_{d - 1}$ is the
area of a unit $(d - 1)$-sphere. The corresponding spectral energy density
reads
\begin{equation}
  \rho (\nu) = \frac{\mathcal{A}_{d - 1}}{c^d} 
  \frac{\mathfrak{g}_{\gamma}}{h^d}  \frac{h^{d + 1} \nu^d}{\mathrm{e}^{\beta
  h \nu} - 1} = \frac{\mathcal{A}_{d - 1}}{c^d} h^{d + 1} \nu^d f_{\gamma} .
  \label{eq:app_f0_rho}
\end{equation}
In the low velocity limit, the macroscopic particle density is
\begin{equation}
  n_A = \int \ensuremath{\boldsymbol{\varpi}}_A p_A^{\hat{0}} f_A \approx m_A
  c \int \ensuremath{\boldsymbol{\varpi}}_A f_A .
\end{equation}

In flat spacetime, when the relative motion of massive particles and the system can be
neglected, the rate of change of the number density of upper energy state
particles $A$ is
\begin{equation}
  \frac{\mathrm{d} n_A}{\mathrm{d} t} \approx \nabla_{\mu} N_A^{\mu} = c \int
  \ensuremath{\boldsymbol{\varpi}}_A \mathcal{C}_{\mathrm{rad}} = c \int
  \ensuremath{\boldsymbol{\varpi}}_A \ensuremath{\boldsymbol{\varpi}}_{\gamma}
  \ensuremath{\boldsymbol{\varpi}}_A'  [W_{\mathrm{abs}}  f_{\gamma} f_A'
  - (W_{\mathrm{emi}} + W_{\mathrm{emi}}
  \mathfrak{g}_{\gamma}^{- 1} h^d f_{\gamma}) f_A],
\end{equation}
where we use Eq.~\eqref{eq:divN} in the second step and substitute $f_{\gamma}^{\ast} = 1 +\mathfrak{g}_{\gamma}^{- 1} h^d
f_{\gamma}$ in the last step. Substituting Eq.~\eqref{eq:app_micro_db},
Eq.~\eqref{eq:app_f0_rho}, expressing $\mathrm{d} p_{\gamma}^{\hat{0}} =
\dfrac{h}{c} \mathrm{d} \nu$ and integrating over $\mathrm{d} \Omega_{d - 1}$,
we get
\begin{align}
  \frac{\mathrm{d} n_A}{\mathrm{d} t} & \approx c \int \mathrm{d} \nu
  \hspace{0.17em} \nu^{d - 2}  \frac{h^{d - 1}}{c^{d - 1}}  \frac{1}{m_A c} 
  \left[ \frac{\mathfrak{g}_A}{\mathfrak{g}_A'} 
  \frac{1}{\mathfrak{g}_{\gamma}}  \frac{c^d}{h \nu^d} \rho (\nu) \int
  \ensuremath{\boldsymbol{\varpi}}_A' W_{\mathrm{emi}} n_A'  \right.
  \nonumber\\
  & \left. \quad - \left( \mathcal{A}_{d - 1}  \int \ensuremath{\boldsymbol{\varpi}}_A
  W_{\mathrm{emi}} + \frac{c^d}{\mathfrak{g}_{\gamma} h \nu^d} \rho (\nu) \int
  \ensuremath{\boldsymbol{\varpi}}_A W_{\mathrm{emi}} \right) n_A \right] . 
\end{align}
Comparing term by term with Eq.~\eqref{eq:app_traditional_rate}, the Einstein
coefficients are identified as:
\begin{align}
  A_{A \to A'} & = \nu^{d - 2}  \frac{h^{d - 1}}{c^{d - 1}} 
  \frac{\mathcal{A}_{d - 1}}{m_A}  \int \ensuremath{\boldsymbol{\varpi}}_A
  W_{\mathrm{emi}},  \label{eq:app_A}\\
  B_{A \to A'} & = \nu^{- 2}  \frac{h^{d - 2} c}{m_A} 
  \frac{1}{\mathfrak{g}_{\gamma}}  \int \ensuremath{\boldsymbol{\varpi}}_A
  W_{\mathrm{emi}},  \label{eq:app_B_emi}\\
  B_{A' \to A} & = \frac{\mathfrak{g}_A}{\mathfrak{g}_A'} \nu^{- 2} 
  \frac{h^{d - 2} c}{m_A}  \frac{1}{\mathfrak{g}_{\gamma}}  \int
  \ensuremath{\boldsymbol{\varpi}}_A' W_{\mathrm{emi}} .  \label{eq:app_B_abs}
\end{align}

Taking the ratio of $A_{A \to A'}$ to $B_{A \to A'}$ recovers the generalized
Einstein relation
\begin{equation}
  \frac{A_{A \to A'}}{B_{A \to A'}} = \frac{\mathfrak{g}_{\gamma}
  \mathcal{A}_{d - 1} h \nu^d}{c^d} . \label{eq:app_einstein_ratio}
\end{equation}
For $d = 3$, with $\mathfrak{g}_{\gamma} = 2$ and $\mathcal{A}_2 = 4 \pi$,
Eq.~\eqref{eq:app_einstein_ratio} reduces to the standard result $\dfrac{8 \pi
h \nu^3}{c^3}$.
The transition rate includes momentum conservation delta functions
\begin{equation}
  W_{\mathrm{emi}} = w_{\ensuremath{\operatorname{emi}}} \delta^{(d + 1)}
  \left( p_A^{\mu} - {p_A'}^{\mu} - p_{\gamma}^{\mu} \right) =
  w_{\ensuremath{\operatorname{emi}}} \delta^{(d)} \left( p_A^{\hat{i}} -
  {p_A'}^{\hat{i}} - p_{\gamma}^{\hat{i}} \right) \delta  \left( p_A^{\hat{0}}
  - {p_A'}^{\hat{0}} - p_{\gamma}^{\hat{0}} \right) .
\end{equation}
Integrating over the phase-space measure $\ensuremath{\boldsymbol{\varpi}}_A =
\dfrac{\mathrm{d}^d p_A}{| (p_A)_{\hat{0}} |}$ yields

\begin{align}
  \int \ensuremath{\boldsymbol{\varpi}}_A W_{\mathrm{emi}} & = \int
  \frac{\mathrm{d}^d p_A}{| (p_A)_{\hat{0}} |}
  w_{\ensuremath{\operatorname{emi}}} \delta^{(d)} \left(  p_A^{\hat{i}} -
  {p_A'}^{\hat{i}} - p_{\gamma}^{\hat{i}} \right) \delta  \left( p_A^{\hat{0}}
  -{p_A'}^{\hat{0}}  - p_{\gamma}^{\hat{0}} \right) \nonumber\\
  & = \left. \frac{w_{\mathrm{emi}}}{p_A^{\hat{0}}} \right|_{p_A^{\hat{i}} =
  {p_A'}^{\hat{i}} - p_{\gamma}^{\hat{i}}} \delta  \left( p_A^{\hat{0}} \left( {p_A'}^{\hat{i}} - p_{\gamma}^{\hat{i}} \right) -
  {p_A'}^{\hat{0}} -
  p_{\gamma}^{\hat{0}} \right) . 
\end{align}
In the low velocity limit, $p_A^{\hat{0}} \approx {p_A'}^{\hat{0}} \approx m_A
c$, making the integration measure symmetric under the interchange of states
$A$ and $A'$ 
\begin{equation}
\int \ensuremath{\boldsymbol{\varpi}}_A W_{\mathrm{emi}} = \int
\ensuremath{\boldsymbol{\varpi}}_A' W_{\mathrm{emi}}.
\end{equation} 
This confirms the relation
\begin{equation}
  \frac{B_{A \to A'}}{B_{A' \to A}} =
  \frac{\mathfrak{g}_A'}{\mathfrak{g}_A} .
\end{equation}
\end{appendices}

\end{document}